\documentclass[iop]{emulateapj}

\usepackage{epsfig}
\usepackage{graphicx}

\newcommand{\kms}{\ifmmode {\rm km\,s}^{-1} \else km\,s$^{-1}$\fi}

\shorttitle{44 GH\MakeLowercase{z} CH$_{3}$OH Maser survey in the GC }
\shortauthors{McEwen et al.}

\begin{document}

\title{44 GH\MakeLowercase{z} Class I Methanol (CH$_{3}$OH) Maser Survey in the Galactic Center}

\author{Bridget C. McEwen}
\affil{The Department of Physics and Astronomy, The University of New Mexico, Albuquerque, NM, 87131}

\author{Lor\'{a}nt O. Sjouwerman}
\affil{National Radio Astronomy Observatory, P.O. Box O, 1003 Lopezville Rd., Socorro, NM, 87801}

\author{Ylva M. Pihlstr\"om\footnote{Y.\ M.\
    Pihlstr\"om is also an Adjunct Astronomer at the National Radio
    Astronomy Observatory}}
\affil{The Department of Physics and Astronomy, The University of New Mexico, Albuquerque, NM, 87131}

\begin{abstract}

We report on a large 44 GHz ($7_0-6_1$ A$^+$) methanol (CH$_3$OH) maser survey of the Galactic Center (GC).  The Karl G. Jansky Very Large Array was used to search for CH$_3$OH maser emission covering a large fraction of the region around Sgr A. In 25 pointings, over 300 CH$_3$OH maser sources ($>10\sigma$) were detected.  The majority of the maser sources have a single peak emission spectrum with line of sight velocities that range from about $-$13 km\,s$^{-1}$ to 72 km\,s$^{-1}$.  Most maser sources were found to have velocities around 35$-$55 km\,s$^{-1}$, closely following velocities of neighboring interacting molecular clouds.  The full width half maximum of each individual spectral feature is very narrow ($\sim$0.85 km\,s$^{-1}$ on average).   In the north, where Sgr A East is known to be interacting with the 50 km\,s$^{-1}$ molecular cloud, more than 100 44 GHz CH$_3$OH masers were detected.  In addition, three other distinct concentrations of masers were found, which appear to be located closer to the interior of the interacting molecular clouds.  Possibly a subset of masers are associated with star formation, although conclusive evidence is lacking.

\end{abstract}
\keywords{masers $-$ ISM: supernova remnants $-$ ISM: individual objects (Sgr A East) $-$ masers $-$ radio lines: ISM }

\section{Introduction}\label{intro}

Astronomical masers form under specific physical conditions and are useful to trace different environments in the interstellar medium (ISM).   Interstellar masers are often found in environments such as star forming regions (SFRs) and supernova remnants (SNRs).  For example, a detection of a collisionally pumped 1720 MHz hydroxyl (OH) maser has traditionally been used as a tracer of shocked regions produced by the interaction of a SNR with a neighboring molecular cloud (MC)  \citep[e.g.,][]{claussen1997, frail1998, yusef2003}.   Another example are the radiatively pumped Class II  methanol (CH$_3$OH) masers lines, which are typically found near young massive stars.  In addition, collisionally pumped Class I CH$_3$OH masers are typically found associated with SNRs and outflows in SFRs \citep[e.g.,][]{beuther2002, voronkov2006, cyganowski2009, fontani2010, sjouw2010, pihl2014, sanna2015}.  Similar to the 1720 MHz OH masers, the Class I 36 and 44 GHz CH$_3$OH maser transitions have also been detected near shocked regions where SNRs are known to be interacting with MCs  \citep[e.g.,][]{sjouw2010, pihl2011, pihl2014}.  

Recent modeling of Class I CH$_3$OH masers in a SNR environment shows that optimal masing conditions for the 44 GHz transition are temperatures $\ge50$ K and densities between $10^4-10^6$ cm$^{-3}$.  Similar temperatures but slightly higher densities in the range of $10^5-10^7$ cm$^{-3}$ are the optimal masing conditions for the 36 GHz transition \citep{mcewen2014, nesterenok2016}.  Because of the large overlap in conditions, these transitions can be found co-spatially, but brighter 36 GHz CH$_3$OH masers are expected to trace higher density regions (e.g., near the actual shock front in SNR/MC interaction regions). This has been supported by observations of bright 36 GHz CH$_3$OH masers lining a known shock front in Sgr A East \citep{sjouw2010,pihl2011}.  

The Sgr A East SNR is located within the inner 12 pc of our complex Galactic Center (GC) and is known to be interacting with two different giant MCs, M$-$0.02$-$0.07 (a.k.a.\,the 50 km\,s$^{-1}$ cloud) and M$-$0.13$-$0.08 (a.k.a.\,the 20 km\,s$^{-1}$ cloud)  \citep[e.g.,][]{mezger1989, coilho2000, amo2011}.  The details of gas transportation in and around the GC are not well understood but different molecular line observations help uncover interaction regions between different environments.  For example, the 50 and 20 km s$^{-1}$ clouds seem to be connected through a 'Molecular Ridge', as suggested by an extensive NH$_3$ study from \citet{coilho2000}, as well as from single dish 36 GHz CH$_3$OH observations \citep{szcz1991}.   In addition, observations by \citet{mcgary2001} suggest possible connections between the MCs and the Circumnuclear Disk (CND) via streamers.  From NH$_3$ observations by \citet{coilho2000} it was deduced that, along the line of sight, Sgr A East is pushing this 50 km s$^{-1}$ MC eastward and away from us (behind),  whereas it also expands into the southern 20 km s$^{-1}$ MC at the same time.  The southern region of Sgr A East is also interacting with the SNR G359$-$0.09 \citep{coilho2000,sjouw2008}. 

These interactions between Sgr A East and its surrounding environment produce collisionally compressed regions of gas some of which are found associated with Class I 36 and 44 GHz CH$_3$OH, as well as collisionally pumped 1720 MHz OH maser emission  \citep[e.g.,][]{yusef1996, sjouw2010, pihl2011}.  It is often speculated that supersonic motions from expanding SNRs may trigger star formation (SF) in neighboring MCs  \citep[e.g.,][]{reynoso2001, cich2014}.  Details of the triggering process and what conditions are necessary have never been clearly outlined or confirmed, partly due to the complexity of the regions in the inner Galaxy and the present lack of star forming regions detected in the GC.  Different stages of SF can be traced with detections of various maser species, for example, 6.7 GHz, as well as the 44 GHz CH$_3$OH maser line have been found closely associated with HII regions, outflows, and H$_2$O masers (typical tracers of SF) (e.g., \citealt{kurtz2004,moscadelli2007,sanna2010}).  It is not clear whether Class I CH$_3$OH masers trace a specific evolutionary stage of SF.  Therefore, the detection of maser lines near these regions (such as bright Class I CH$_3$OH or 22 GHz H$_2$O lines) may unveil new sites of, and conditions necessary for SF activity in the GC.  In addition, the proper understanding of the physical conditions in these regions may also be important for cosmic ray modeling \citep[e.g.,][]{drury1994,abdo2010,cristofari2013}. 

Combining maser observations of SNRs and their surrounding environments along with modeling of conditions necessary for the formation of CH$_3$OH masers in SNRs, physical conditions of the gas where CH$_3$OH is detected can be constrained \citep{mcewen2014}.  In this context, by studying the distribution of different maser transitions near Sgr A East, we aim to investigate the presence of distinct gas motions in the GC, as well as possibly uncover new sites of star formation.  In this study, we report on an extensive 44 GHz CH$_3$OH maser emission survey of Sgr A East and its surrounding environment taken with the Very Large Array (VLA).   A 36 GHz CH$_3$OH maser survey towards the same region along with a full analysis of the combined 36 and 44 GHz maser detections will be reported on in a future publication.

\section{44~GH\MakeLowercase{z} VLA Observations and Calibration}\label{calib}

We are reporting on the results from Q-band VLA observations (project code S3115) of the SNR Sgr A East and its surrounding environment.   The Q-band B configuration observations were taken on April 20 and 23, 2011 to observe the $J=7_0\rightarrow6_1A^+$ rotational transition of CH$_3$OH at the rest frequency of 44.069 GHz.  Figure\,1 displays the 25 pointing positions overlaid on a 1720 MHz continuum image of the Sgr A region.  The VLA primary beam is $1.02'$ at 44 GHz, with a typical synthesized beam size of $0.38''\times0.19''$. The 25 pointings covered a region of roughly $8'\times6'$.  Although the area is not Nyquist sampled, regions in between different pointings were also searched for potential masers by imaging beyond the primary beam.  These candidate sources have been considered physical if detected in more than one pointing. 

The 44 GHz observations were separated into 256 frequency channels covering 16 MHz of bandwidth.  We sampled a velocity range between $-24$ and 84 km s$^{-1}$ with a resolution of about $0.4$ km s$^{-1}$.   Sgr A*, located in pointing A00, was used for phase calibration and 3C286 was used as a flux and bandpass calibrator. The total on source time for each pointing including both days of observation was about 8 minutes.  Each pointing was individually imaged with 2048$\times$2048 pixels of $0.036''$ for the central 250 channels ($-23.2$ to 82.7 km\,s$^{-1}$).  Typical noise rms values were around 2.9 mJy\,beam$^{-1}$ per channel. 

The data were reduced, calibrated, and imaged using standard procedures in NRAO's Astronomical Image Processing System (AIPS) pertaining to spectral line data.  Fields with bright maser sources ($>1$ Jy\,beam$^{-1}$) were self-calibrated in order to improve the dynamic range of the final maps, and to minimize the number of false detections due to side-lobes.  The peak flux densities were also corrected for primary beam attenuation using the AIPS task PBCOR. 

\begin{figure}
\begin{center}
  \includegraphics[scale=.52]{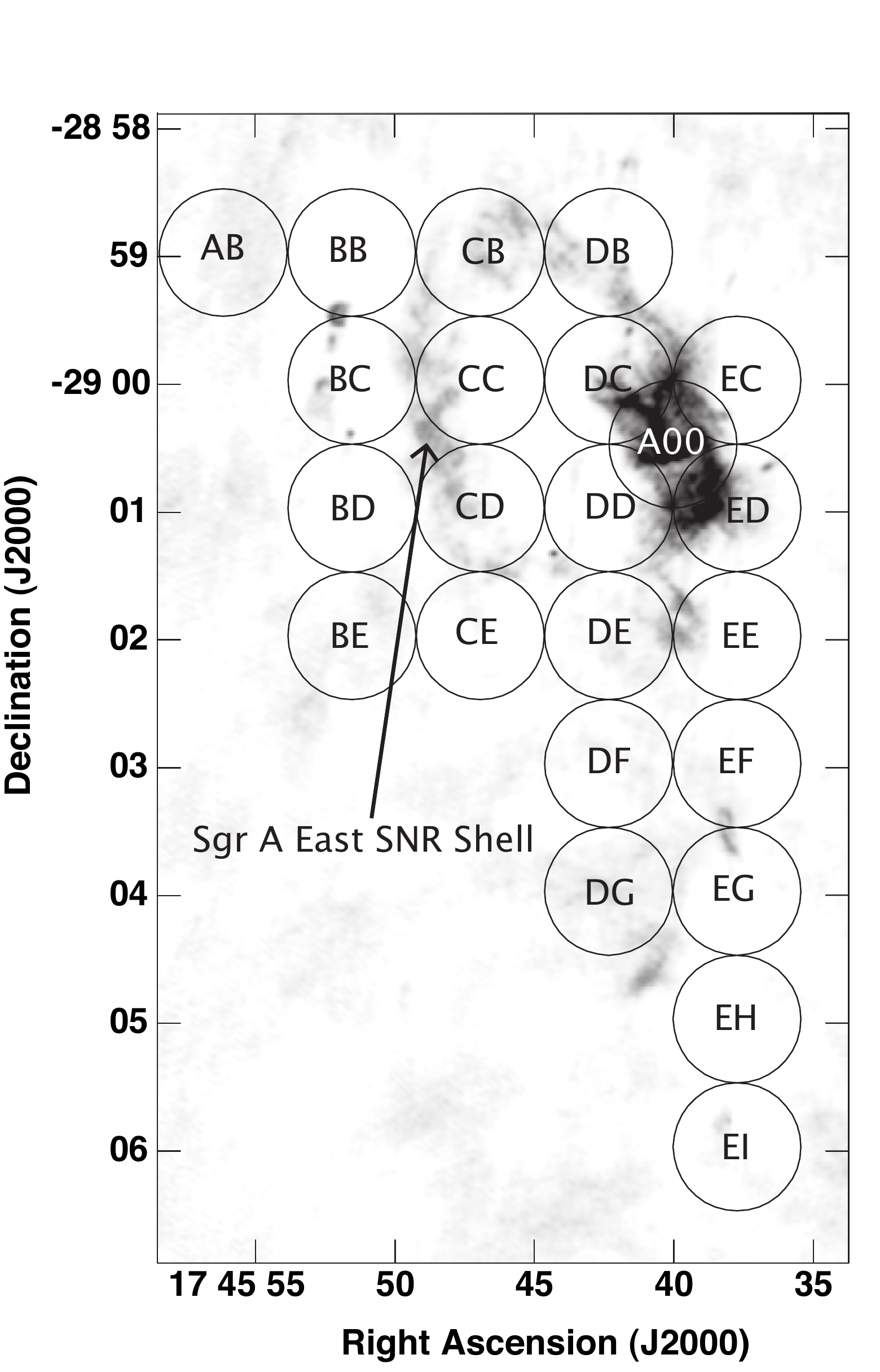}
\caption{44 GHz VLA observing pointing positions towards the Sgr A region.  The ring-like structure pointed at by the arrow outlines the radio continuum emission of the Sgr A East SNR.}
\label{f1}
\end{center}
\end{figure}

\section{Results}\label{results}
\subsection{Identification Method}
Sgr A East is located in a complex and chemically rich environment and many Class I CH$_3$OH masers have been previously detected towards various regions near this SNR \citep{yusef2008, sjouw2010, pihl2011, yusef2013}.  In order to search each pointing efficiently, an automated search method using a variety of AIPS tasks was developed to identify maser candidates with flux densities exceeding 10 times the rms noise.  These candidates were then sorted according to a confidence in the detection, which depended on the size of the emission region and its signal to noise ratio. The highest ranked candidates were then manually inspected to determine if each was actual  emission.  Spectral profiles (flux density versus velocity) were produced for the brightest pixel in each region (Fig.\,7).   

\subsection{Maser Identification}
A total of 318 100\% confident 44 GHz CH$_3$OH emission regions were found in the 25 pointings.  All exceeded a 10$\sigma$ rms noise limit.  The spectral parameters of each emission region are presented in Table 1.  The coordinates of each source can be found in Columns 3 and 4 (with an estimated positional accuracy of $\sim0.5''$), which correspond to the peak brightness, I$_{peak}$ of each emission region (Column 5).  Figure\,2 shows a histogram of the distribution of the  peak flux density values of the maser emission the majority of which are $\lesssim0.8$ Jy\,beam$^{-1}$.  The emission was detected across almost the entire region observed with some regions of high concentration.  A large abundance of maser emission was detected toward the NE and to the SW, as well as a small amount encompassing the SNR.  The maser emission extends to the east about  $2'$ (4.6 pc) from the NE boundary of the SNR shell and about $5' 15''$ (12.1 pc) to the south of the SW boundary of the SNR shell.  The majority of emission is unresolved and the brightness temperatures, T$_b$, listed in Column 8, are lower limits calculated using the half power beam width of the VLA in B configuration.  The minimum brightness temperatures range from $19\times10^2$ to $52\times10^4$ K.  The 44 GHz CH$_3$OH masers in this study that have previously detected counterparts are indicated in Table 1 with [R].

The peak velocities vary across the region observed, ranging from about $-13$ to 72 km\,s$^{-1}$ and are listed in Column 6.    Figure\,3 shows two simple histograms of the velocity of maser emission in the NE and SW regions observed separated by a declination of $-29^{o}$ $02'$ $30'$ (between rows E and F in Fig.\,1).  The color scheme in Fig.\,3 represents the different velocities of the emission. The full width half maximum (FWHM) of the brightest peak is listed in Column 7 and  was estimated from the number of channels at half I$_{peak}$ with an error of half of a channel ($\pm0.2$ km\,s$^{-1}$).  A gaussian fit was inaccurate to estimate the FWHM because most to the peaks are only a couple channels wide.  Instead, we used the number of channels at half peak to approximate an upper limit to the FWHM for each peak.  The FWHM of each emission peak listed in Table 1 is narrow ranging from about 0.4 to 3.0 km\,s$^{-1}$ (1 to 7 channels).  These line-widths have similar values to previously detected 44 and 36 GHz CH$_3$OH masers in Sgr A East and other SNRs \citep{sjouw2010, pihl2011}.  The estimated thermal line-width using the lowest calculated T$_b$ (1,900 K) is about 1.6 km\,s$^{-1}$.   Two sources (119 and 160) have measured FWHMs close to the thermal linewidth.  A gaussian was fit to the spectral profiles of these two sources but the errors in the fits were large, therefore we conclude that 119 and 160 are not thermal sources.   The measured FWHM of the remaining emission regions are less than the calculated thermal line-width for their minimum T$_b$, therefore we conclude that all the peaks listed in Table 1 are maser emission and non-thermal in nature.

Although the majority of masers detected were found to be single peaks of emission, as can be seen in Fig.\,7, some of the spectra show multiple spectral peaks at a single position ($\sim$\%27).  The sources in Table 1 that are noted with a [M] indicate the sources that display multiple spectral features detected at the same position, some with complicated and broad structure in their spectral profiles. Some of the multiple peak features may imply that we are observing partly unresolved structures within the VLA beam.  The emission regions with multiple peaks cover the entire region observed but are mostly concentrated in pointings BB and BC to the NE and EI and EG to the SW.  Given the sensitivity of the VLA and the beam size in B-configuration, our observations may not be sensitive to thermal or extended sources below $4\sigma$ corresponding to a brightness temperature less than $\sim$370 K.

\begin{figure} 
\begin{center}
 \includegraphics[scale=.25]{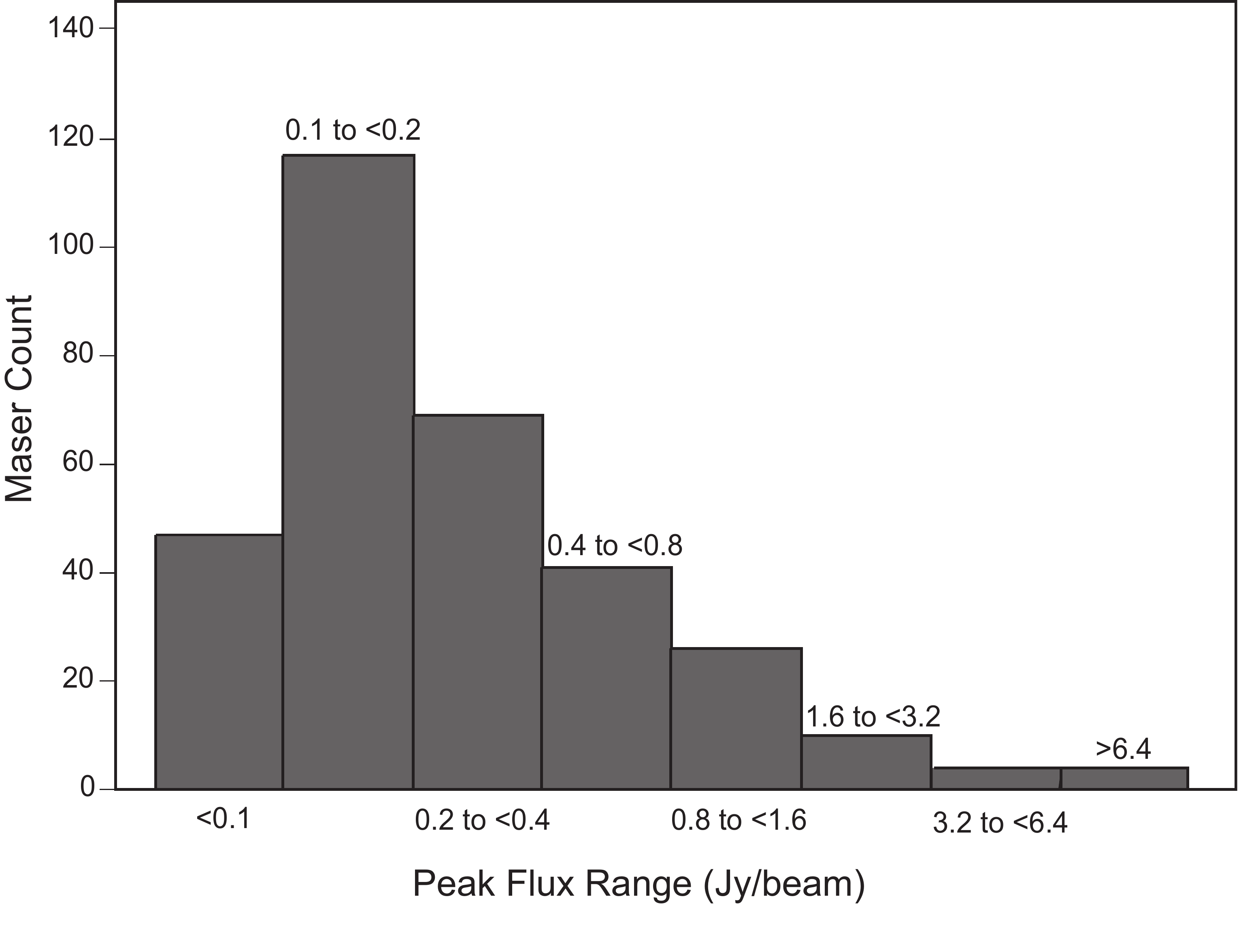}
\caption{A histogram showing the peak flux density distribution of the maser emission.}
\end{center}
\end{figure} 

\begin{figure} 
\begin{center}
 \includegraphics[scale=.43]{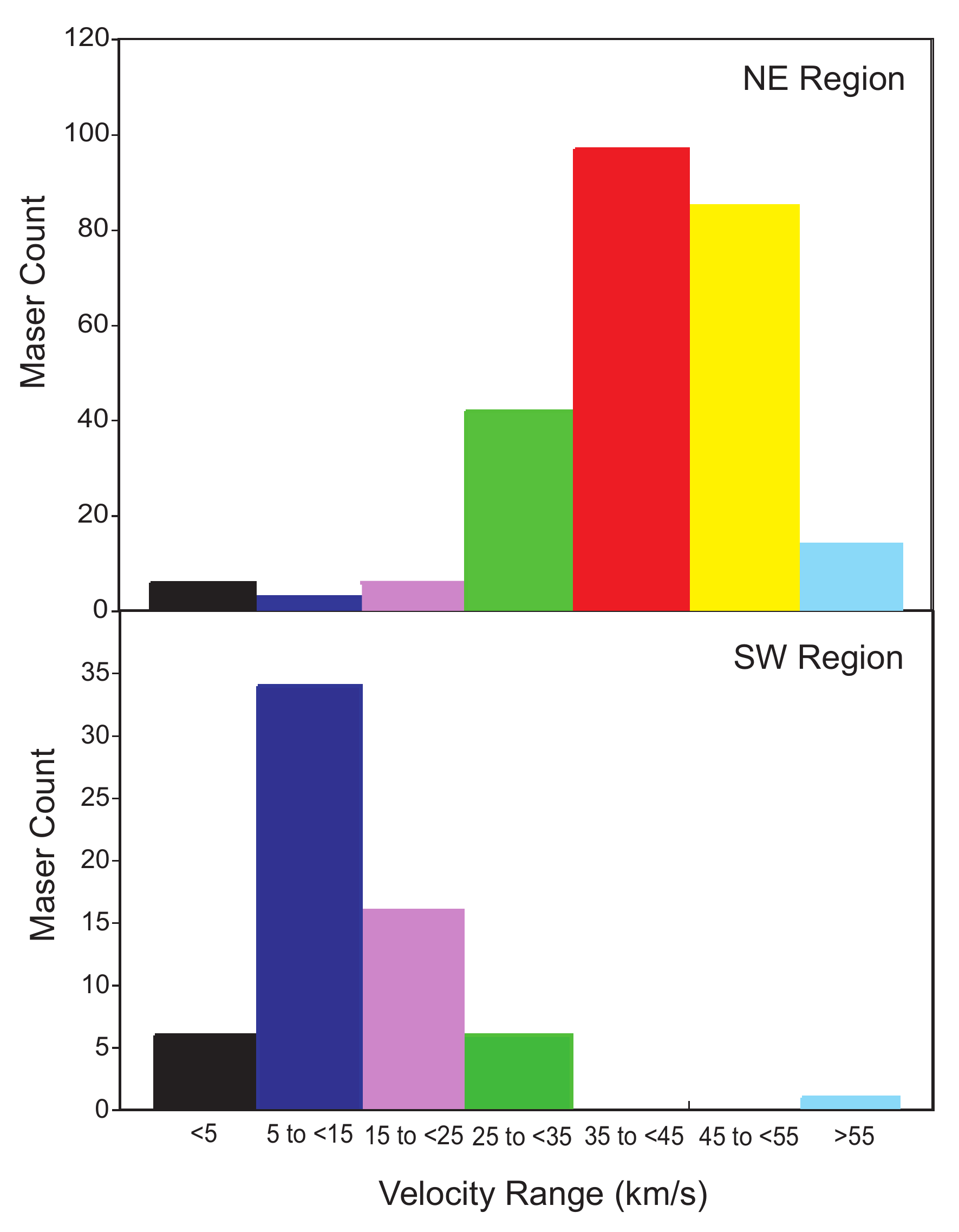}
\caption{Two histograms showing the velocity distribution of the maser emission in the NE (top) and SW (bottom) regions observed. The masers corresponding to these regions are separated by a declination of $-29^{o}$ $02'$ $30''$.  The different colors correspond to the velocity of the masers.  This color scheme is used in later figures to indicate the velocity of each maser along with their position. }
\end{center}
\end{figure}

 \section{Discussion}

Various regions in the Sgr A East environment have previously been searched for different maser transitions.  Four collisionally pumped maser transitions have been found, namely the 22 GHz H$_2$O, 1720 MHz OH, 36 GHz CH$_3$OH, and 44 GHz CH$_3$OH lines \citep{yusef1996,  sjouw2002, yusef2008, sjouw2010, pihl2011, yusef2013}.  The distribution trends of all these maser species in a few regions will be briefly discussed.  

\subsection{CH$_3$OH Maser Distribution}

Figure\,4 shows the positions of the 44 GHz CH$_3$OH maser sources (crosses) detected from this survey within the observed region indicated by the black dashed line, overlaid on a 1720 MHz continuum image. Previously detected 36 GHz (circles) and 44 GHz (triangles) CH$_3$OH, 1720 MHz OH (squares), and 22 GHz H$_2$O (plus signs) masers are also plotted \citep{yusef1995, yusef1996,  sjouw2002, yusef2008, sjouw2010, pihl2011, yusef2013}.  The color of each symbol represents the velocity bin of the maser according to the scheme in Fig.\,3, for example, light blue symbols represent velocities $>55$ km\,s$^{-1}$.  Typical beam sizes and channel rms values from the previous observations are as follows: $15''$ and 15 mJy\,beam$^{-1}$ \citep{yusef1996}, $2.5''\times1.9''$ and 15 mJy\,beam$^{-1}$ \citep{sjouw2002}, $5.8''\times3.9''$ and 14.1mJy\,beam$^{-1}$ \citep{pihl2008}, $0.2-0.4''$ and 10-12 mJy\,beam$^{-1}$ \citep{sjouw2010}, $1.3''\times0.5''$ and 15-20 mJy\,beam$^{-1}$ \citep{pihl2011}, and $1.8''\times0.7''$ and 2.5 mJy\,beam$^{-1}$ \citep{yusef2013}. The previously detected 44 GHz CH$_3$OH masers from \citet{pihl2011} and \citet{yusef2008} were also detected in this survey and consistent within our positional accuracy, and have velocities within $\pm2$ km\,s$^{-1}$ of our listed V$_{peak}$.   Figure\,4 shows that the majority of emission has velocities around 10 km\,s$^{-1}$ to the SW region and around $45$ km\,s$^{-1}$ to the NE region of the Sgr A East shell.

The most noteworthy result from this survey is the large number of CH$_3$OH masers detected within the inner parsecs of our GC.  The CH$_3$OH maser emission detected in other Galactic SNRs interacting with MCs pales in comparison to Sgr A East.  For example, in a targeted search towards the SNR W28, few 36 and 44 GHz CH$_3$OH masers were found.  Towards the SNR G1.4$-$0.1, which is interacting with at least 2 MCs, only 36 GHz CH$_3$OH maser emission was detected, and none at 44 GHz \citep{pihl2014}.  It is widely accepted that CH$_3$OH forms on the surface of icy dust grains and then is released into the gas-phase via some heating mechanism, for  example, through UV radiation, shocks from cloud-cloud interactions, SNR-cloud interactions, by expanding HII regions, and by young and old stellar outflows \citep[e.g.,][]{garrod2008, whittet2011, ruiz2016, yusef2013}.  This enhancement of CH$_3$OH detected near Sgr A East may not be surprising because the GC is extremely chemically rich and subject to shocks.  In addition, the enhancement of CH$_3$OH may be driven by cosmic ray irradiation, more so than in other regions of the Galaxy where the cosmic ray ionization rate is lower \citep{yusef2013,pihl2014,mills2015}.

\begin{figure*} 
\begin{center}
 \includegraphics[scale=1.7]{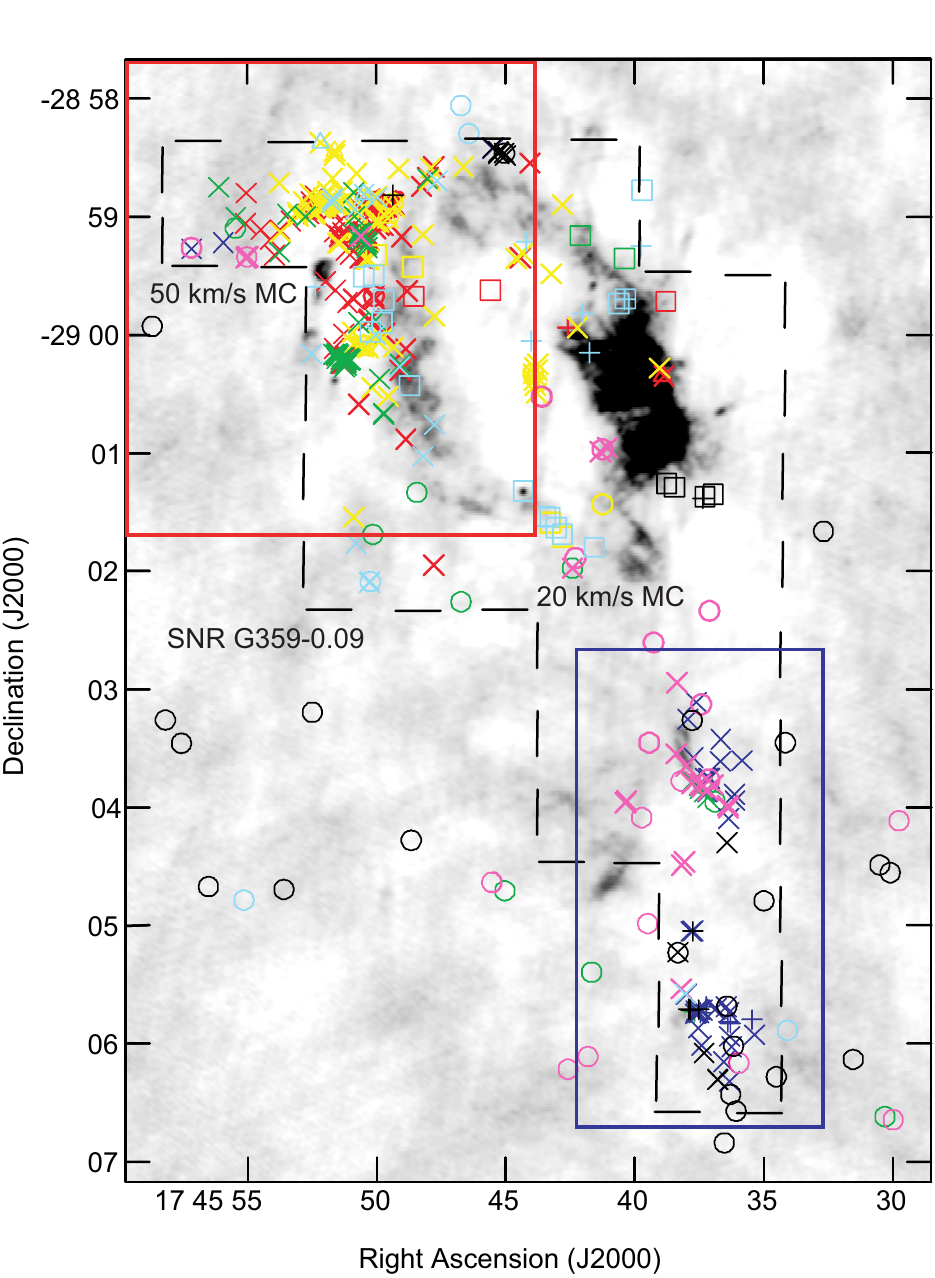}
\caption{Positions of the 44 GHz CH$_3$OH masers (crosses) overlaid on a 1720 MHz continuum emission of the Sgr A East environment.  In addition, previously detected 44 GHz CH$_3$OH (triangles), 36 GHz CH$_3$OH (circles), 22 GHz H$_2$O (plus-signs), and 1720 MHz OH (squares), masers are overlaid, for details see \citet{yusef1995,yusef1996,sjouw2002,yusef2008,sjouw2010,pihl2011,yusef2013}.  The color of each symbol represents the velocity of each maser according to the scheme in Fig.\,3.  The dashed black line indicates the the observed region in this study.  The red and blue boxed regions correspond to the enlarged NE and SW regions in Fig.\,5 and 6, respectively.}
\end{center}
\end{figure*}

\subsubsection {NE Region}
A high concentration of masers is found in the NE region of Sgr A East, where this SNR is interacting with the 50 km\,s$^{-1}$ MC and borders the radio continuum shell of Sgr A East, as can be seen in Fig.\, 5 (zoom of the red, boxed region in Fig.\,4).  The vast majority of the 44 GHz masers in this clump are found to have velocities similar to that of the MC, around 50 km\,s$^{-1}$ or less (red and yellow symbols).   Many of these masers are coincident with a known shock front \citep{pihl2011} and they seem to follow a sharp boundary along the edge of the SNR radio continuum emission, outlined by the black rectangle in Fig.\,5.   The brightest 44 GHz CH$_3$OH masers detected in this region, with $I_{peak}$ between 5.87 and 16.16 Jy\,beam$^{-1}$, are significantly weaker than the brightest 36 GHz CH$_3$OH masers detected in the same region.  The lowest flux density ratios between these two maser species in this region is $\sim$5. Based on modeling results from \citet{mcewen2014}, this implies a high density region ($n>10^6$ cm$^{-3}$).   In addition, the high concentration of masers in this region is spatially coincident with strong SiO ($2-1$) emission (between 20 and 50 km\,s$^{-1}$), as can be seen in \citet{yusef2013}, which is also indicative of high density shocked gas. Very few masers are detected to the west (right) of this boundary, which is in agreement with what was seen by \citet{pihl2011}.  This means that the physical conditions to the west of the shock front are not conducive for CH$_3$OH maser emission.  It is possible that there may be a lower abundance of CH$_3$OH in this region, due to UV photodissociation of the CH$_3$OH molecule as was speculated by \citet{yusef2013}. 

\begin{figure*} 
\begin{center}
 \includegraphics[scale=3.5]{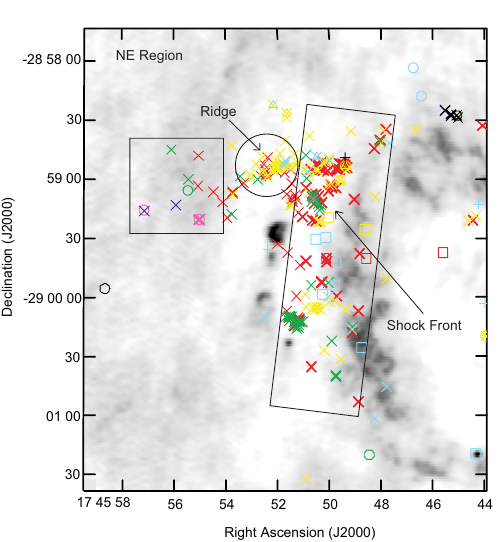}
\caption{Positions of the 44 GHz CH$_3$OH masers (crosses) overlaid on a 1720 MHz continuum emission of the zoomed in NE region of the Sgr A East environment indicated by the red box in Fig.\,4.  In addition, previously detected 44 GHz CH$_3$OH (triangles), 36 GHz CH$_3$OH (circles), 22 GHz H$_2$O (plus-signs), and 1720 MHz OH (squares), masers are overlaid, for details see \citet{yusef1995,yusef1996,sjouw2002,yusef2008, sjouw2010,pihl2011,yusef2013}.  The color of each symbol represents the velocity of each maser according to the scheme in Fig.\,3.  The black rectangle indicates a large abundance of masers in a region where the 50 km\,s$^{-1}$ MC boarders the radio continuum shell.  The black circle outlines a cluster of 44 GHz CH$_3$OH maser emission forming a dense ridge.}
\end{center}
\end{figure*}

The collisionally excited 1720 MHz OH masers form under similar physical conditions compared to Class I CH$_3$OH, but are found offset from the CH$_3$OH maser positions.  This means they are most likely formed in different regions in the shocked gas \citep{pihl2008, pihl2014, mcewen2014}.  The majority of the 44 GHz masers in the NE region have slightly lower velocity spread compared to the OH masers, which have an average of $\sim57$ km\,s$^{-1}$, and are located just to the SW of the group of 44 GHz CH$_3$OH masers (in the rectangle region).  This implies that the OH masers are located in a region of the shock that is more turbulent and disturbed, namely in the post shocked gas.  This strengthens the same conclusion that was drawn based on previous but more sparse observations of a CH$_3$OH maser survey by \citet{pihl2011} and reinforces the idea that bright 36 GHz CH$_3$OH masers coincident with weaker 44 GHz CH$_3$OH masers detected in a SNR/MC interaction region trace a region closer to the actual shock front compared to OH maser emission.  

Just to the east of this front,  there appears to be another cluster of  44 GHz CH$_3$OH maser emission forming a dense ridge, outlined by the black circle in Fig.\,5.   Given the distance to the GC of 8.5 kpc, this ridge is about 1.2 pc to the east of the shock front.  As suggested by \citet{mcewen2014}, these masers may be associated with a possible newly detected young SNR embedded in the 50 km\,s$^{-1}$ MC.  In this region, \citet{tsuboi2011, tsuboi2012} detected a dense shocked molecular shell region based on CS (J$=1-0$) observations and high SiO/H$^{13}$CO$^+$ ratios.  The SiO emission ranges from about 15 to 45 km\,s$^{-1}$, which agrees with the velocity of the masers.  Alternatively, they could be excited by an internal shock in the core of the cloud generated by star formation (SF).

\subsubsection{SW Region} 
Two distinct concentrations of masers are found to the SW of Sgr A East, where this SNR is interacting with the 20 km\,s$^{-1}$ MC (Fig.\, 6). Here, the vast majority of the 44 GHz masers have velocities that range from 5 to 15 km\,s$^{-1}$ (blue symbols).  However, in the northern cluster (outlined by the black square in Fig.\,6) several masers are found to have velocities similar to that of the MC, closer to 20 km\,s$^{-1}$ (pink and green symbols) and outline a known non-thermal filament in this region (SgrA-F) \citep{ho1985}.  In the southern cluster (outlined by the black circle) several masers have velocities less than 5 km\,s$^{-1}$ (black symbols), indicating a different origin.  The masers in the southern cluster form a rough circle centered around a declination of $-29^o$ $05'$ $55''$ and is located about $10''$ to the SW of a known HII region (SgrA-G, 17h 45m 38.21s $-29^o$ $05'$ $45.5''$) \citep{ho1985}. These two distinct clusters can also be seen in SiO (2-1) with comparable velocities around 20 km\,s$^{-1}$ (northern) and 0 km\,s$^{-1}$ (southern), in good agreement with the velocities of other masers in these regions  \citep{tsuboi2011}.   The majority of the 36 GHz masers previously detected are around 17 km\,s$^{-1}$ (northern cluster) and around 0 km\,s$^{-1}$ (southern cluster).  Given the distance of 8.5\,kpc to the GC, the separation between these two clumps is about 5.8 pc.  The scale of both of these clusters is roughly the same, $\sim1.6$ pc, which is about double the size of a typical compact HII region \citep{kurtz2005}.  No 1720 MHz OH masers are detected near these two clusters.

\begin{figure*} 
\begin{center}
 \includegraphics[scale=4]{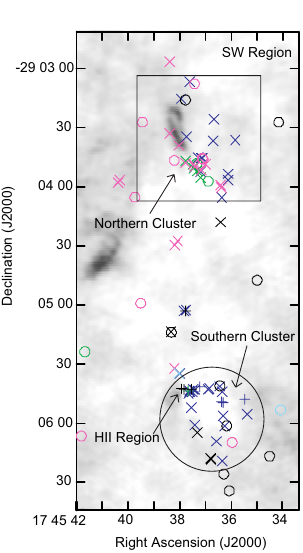}
\caption{Positions of the 44 GHz CH$_3$OH masers (crosses) overlaid on a 1720 MHz continuum emission of the zoomed in SW region near Sgr A East indicated by the blue box in Fig.\,4.  In addition, previously detected 44 GHz CH$_3$OH (triangles), 36 GHz CH$_3$OH (circles), 22 GHz H$_2$O (plus-signs), and 1720 MHz OH (squares), masers are overlaid, for details see \citet{yusef1995,yusef1996,sjouw2002,yusef2008,sjouw2010,pihl2011,yusef2013}.}
\end{center}
\end{figure*}

 \subsection{Possible Star Formation Near Sgr A East}
  
 Observationally, 44 GHz CH$_3$OH lines are found to be more common and brighter near SF regions compared to the 36 GHz line, although both transitions have been found co-located (e.g., \citealt{voronkov2014}).  It is not clear whether the Class I CH$_3$OH maser traces a specific evolutionary stage of SF.  \citet{voronkov2014} suggest that Class I  CH$_3$OH masers may be found at multiple epochs throughout the evolution of a massive star. They find Class I CH$_3$OH masers near regions with other maser sources that trace early evolutionary stages (e.g., 6.67 GHz CH$_3$OH), as well as older stages (e.g., OH).   Another possibility may be that Class I CH$_3$OH masers arise from/in different types of shocks in SF regions, including young and old outflows, cloud/cloud collisions, and expanding HII regions \citep{ruiz2016}. 

\citet{mcewen2014} show that in a SNR environment (low dust temperature and IR radiation), given a specific CH$_3$OH abundance, the 36 GHz maser line dominates at higher number densities and the 44 GHz line at lower densities.  These transitions are both collisionally pumped lines, but at lower densities where the 44 GHz line dominates, it is possible that IR pumping may play an important role.  If so, it could help explain why the Class I 44 GHz line is more common in SF regions, where IR radiation and dust emission are more prevalent.  In fact, modeling by \citet{nesterenok2016} shows that strong radiation fields can quench other collisionally pumped Class I CH$_3$OH lines (e.g., the 25 GHz line), while the radiatively pumped Class II lines (e.g., the 6.7 GHz line) become brighter.  However, it is also found that both Class I and Class II masers can be bright and exist in the same region with high IR radiation fields.  The modeling work carried out by \citet{nesterenok2016} did not extend to the 36 and 44 GHz lines.  Using the online radiative transfer modeling program (RADEX,  van der Tak et al. 2007), taking into account a strong external radiation field (100K) appropriate for SF regions, we indeed find conditions where the 44 GHz and 36 GHz lines simultaneously exist.  We also find conditions where the 44 GHz line dominates over the 36 GHz line (e.g., high temperatures $\sim$200 K and densities $\sim$10$^4$ cm$^{-3}$).  This supports the idea that strong IR radiation fields could influence the production of Class I CH$_3$OH lines.

 Based on our VLA observations, three regions that are offset from the radio SNR shell can be identified, which will be discussed for possible star formation association.  One region is to the NE of the SNR, seemingly located towards interior of the 50 km s$^{-1}$ MC, Fig.\,5. The other two are to the SW of the SNR near the southern cluster of masers labeled in Fig.\,6. 
 
{\bf Region one:}  Towards the far NE region outlined by the black square in Fig.\,5, there are three recently detected 36 GHz CH$_3$OH masers \citep{yusef2013}.  These masers are located more towards the interior of the 50 km\,s$^{-1}$ MC.  Two of the masers are spatially coincident with a few 44 GHz masers with similar velocities.  In addition, these masers have slightly lower velocities compared to those towards the edge of the Sgr A East shell to the west.  Despite searches, radio continuum and H$_2$O masers have not been detected in this NE region, which are often signposts of HII regions. 

{\bf Region two:} In the SW region (Fig.\,6) towards the core of the 20 km\,s$^{-1}$ MC interaction, there is a known HII region (SgrA-G) just to the NW of the the southern cluster where H$_2$O masers are detected \citep{sjouw2002} that have similar velocities to the CH$_3$OH masers.  The fact that there are both H$_2$O masers and 44 GHz CH$_3$OH masers in this region suggest possible star formation.  It is probable that some of the CH$_3$OH masers closer to the HII region are tracing outflows or shocks produced from the expanding HII region.  

{\bf Region three:}  In the SW southern cluster (Fig.\,6), the circular distribution nature of the CH$_3$OH masers could mean that they are tracing outflows from an undetected SF region (Fig.\,6).  In addition, no 1720 MHz OH maser emission has been detected in this region, which would be consistent with an early stage of SF.  Additional observations of other SF tracers, will hopefully shed light on the properties of this region.

Note that the four known compact HII regions (A-D; Ho et al. 1985) located just to the east of the region outlined by the black rectangle in Fig.\,5 are located in the foreground and therefore are not related to the shock front of Sgr A East \citep{sjouw2008}.  Although spatially coincident, the few 44 GHz CH$_3$OH masers and one H$_2$O maser are not associated with these HII regions. 

\section{Conclusions}

Over 300 masers were detected in the Sgr A East region at 44 GHz.  The majority of the maser emission is found to be associated with the interaction of the SNR with the neighboring MCs to the NE and SW of the SNR.  We summarize the results of this survey in three main points:  first, the distribution and abundance of 44 GHz CH$_3$OH masers is very different compared to the 1720 MHz OH masers.  The 44 GHz CH$_3$OH masers are much more abundant than OH and are not found co-spatial with the 1720 MHz OH masers, which suggests they are sustained in different regions of the shocked environment.  Second, the brightest 44 GHz CH$_3$OH masers detected in this study are significantly weaker compared to the brightest 36 GHz CH$_3$OH masers detected in the same region, located in the NE shocked region of Sgr A East, which indicates a high density.  Third, it is possible that some of the masers are tracing sites of star formation, although conclusive evidence does not exist at this time.  A more complete survey of 36 GHz maser emission is underway, and will be used to complete a full analysis of this region.

\acknowledgments We thank NASA for support
under FERMI grant NNX12AO77G. B.C.M.\, acknowledges support from
the NM Space Grant Consortium under the Graduate Research Fellowship program.
The National Radio Astronomy Observatory is a facility of the National
Science Foundation operated under cooperative agreement by Associated
Universities, Inc.

\LongTables
\begin{longtable*}{rlcccrrrc}
\caption{44 GHz CH$_3$OH Detections}\\ 
\hline\hline
\# & Pntg & RA & DEC & I$_{peak}$ & V$_{peak}$ & FWHM & min T$_b$ &Notes \\ 
     &          & (J2000) & (J2000) & (Jy beam$^{-1}$) & (km s$^{-1}$) & (km s$^{-1}$) & ($10^2$ K)  \\
\hline 
\endfirsthead

\multicolumn{8}{@{}c}{44 GHz Detections continued \ldots}\\

\hline
\# & Pntg & RA & DEC & I$_{peak}$ & V$_{peak}$ & FWHM & min T$_b$& Notes \\ 
     &          & (J2000) & (J2000) & (Jy beam$^{-1}$) & (km s$^{-1}$) & (km s$^{-1}$) & ($10^2$ K)   \\
\hline 
\endhead
\hline
\endfoot
\hline
\endlastfoot

1&EI&17 45 35.36 &$-$29 05 55.5&0.08&$+$9.2&0.4&         26  \\
2&EG&17 45 35.84 &$-$29 03 36.3&0.12&$+$6.2&0.8&        39    &M\\
3&EG&17 45 36.12 &$-$29 03 56.6&0.07&$+$11.3&0.4&     23     &M\\
4&EG&17 45 36.13 &$-$29 03 53.4&0.25&$+$11.3&0.4 &         81   \\
5&EI&17 45 36.24 &$-$29 06 01.6&0.38&$+$7.5&0.4&        120   \\
6&EI&17 45 36.32 &$-$29 05 56.4&0.06&$+$6.6&0.4&         19   \\
7&EI&17 45 36.33 &$-$29 06 19.2&0.18&$+$10.9&0.4&         58 \\
8&EI&17 45 36.34 &$-$29 05 44.2&0.06&$+$9.6&0.8&    19     \\

9&EG&17 45 36.36 &$-$29 04 05.6&0.08&$+$6.6&0.4&         26   \\
10&EG&17 45 36.37 &$-$29 04 00.4&0.20&$+$16.8&0.4&        65  \\
11&EG&17 45 36.40 &$-$29 03 59.2&0.09&$+$20.6&0.4&             29\\
12&EG&17 45 36.42 &$-$29 04 18.0&0.09&$+$1.5&0.4&         29   \\
13&EI&17 45 36.45 &$-$29 05 41.4&0.09&$+$6.6&0.8&           29    \\
14&EI&17 45 36.56 &$-$29 06 09.3&0.09&$+$9.2&0.4&           29     \\

15&EG&17 45 36.66 &$-$29 03 25.6&0.10&$+$6.6&0.4&        32\\
16&EG&17 45 36.69 &$-$29 03 36.7&0.13&$+$6.6&0.4&            42 \\
17&EG&17 45 37.11 &$-$29 03 45.4&0.25&$+$13.8&0.4&       81    &M\\
18&EI&17 45 36.78 &$-$29 06 18.4&1.25&$-$4.4&0.4&      400  \\
19&EI&17 45 36.79 &$-$29 06 18.1&1.22&$-$5.7&2.8&         390         &         M\\
20&EI&17 45 36.79 &$-$29 06 18.1&0.68&$-$4.9&1.6&            220    &         M\\
21&EI&17 45 36.84 &$-$29 05 42.3&1.17&$+$13.0&0.8&      380  \\

22&EI&17 45 36.89 &$-$29 05 42.9&0.11&$+$11.7&0.4&          36\\
23&EG&17 45 36.98 &$-$29 03 48.6&0.17&$+$21.9&0.4&        55    \\
24&EG&17 45 37.12 &$-$29 03 45.4&0.44&$+$14.3&0.4&       140     \\
25&EG&17 45 37.15 &$-$29 03 45.4&0.22&$+$14.3&1.6&              71     &         M\\
26&EG&17 45 37.16 &$-$29 03 51.4&0.08&$+$15.5&0.4&          26 \\
27&EG&17 45 37.17 &$-$29 03 55.2&0.11&$+$27.9&0.4&    36    \\

28&EG&17 45 37.18 &$-$29 03 52.3&0.09&$+$14.7&0.4&            29  \\
29&EG&17 45 37.27 &$-$29 03 45.2&0.11&$+$12.1&0.4&      36    \\
30&EG&17 45 37.30 &$-$29 03 52.1&0.17&$+$29.6&0.8&            55   \\
31&EI&17 45 37.31 &$-$29 06 04.8&0.07&$-$2.7&0.8&        23    \\
32&EI&17 45 37.40 &$-$29 05 43.1&0.15&$+$10.9&0.4&          48  \\
33&EG&17 45 37.40 &$-$29 03 49.8&0.08&$+$24.9&0.4&         26    \\
34&EI&17 45 37.41 &$-$29 06 01.0&0.16&$+$7.5&0.4&        52  \\

35&EG&17 45 37.42 &$-$29 03 49.4&0.07&$+$25.7&0.4&      23     \\
36&EG&17 45 37.46 &$-$29 03 48.4&0.07&$+$23.6&0.4&        23    \\
37&EG&17 45 37.46 &$-$29 03 50.5&0.54&$+$25.7&0.4&     170     \\
38&EI&17 45 37.47 &$-$29 05 42.9&0.42&$+$11.3&0.4&       140 \\
39&EG&17 45 37.49 &$-$29 03 48.7&2.85&$+$25.3&0.4&       920   &         M\\
40&EI&17 45 37.50 &$-$29 05 44.3&2.12&$+$12.1&0.4&        680    &         M\\
41&EI&17 45 37.54 &$-$29 05 52.2&0.30&$+$6.2&0.8&     97   \\
42&EI&17 45 37.60 &$-$29 05 44.2&0.17&$+$12.6&0.4&     55     \\
43&EF&17 45 37.61 &$-$29 03 06.7&0.10&$+$6.2&0.4&         32  \\

44&EI&17 45 37.62 &$-$29 05 43.5&0.63&$+$13.8&0.4&      200   \\
45&EI&17 45 37.65 &$-$29 05 45.0&0.20&$+$14.7&0.4&        65    \\
46&EG&17 45 37.73 &$-$29 03 34.7&0.06&$+$6.2&0.4&              19   &         M\\
47&EH&17 45 37.75 &$-$29 05 02.8&0.77&$+$13.8&0.4&     250   \\
48&EG&17 45 37.76 &$-$29 03 46.8&0.22&$+$26.6&0.4&            71   \\
49&EG&17 45 37.77 &$-$29 03 47.9&0.25&$+$24.5&0.4&       81  \\
50&EH&17 45 37.82 &$-$29 05 03.1&0.38&$+$13.4&0.4&     120     \\

51&EF&17 45 37.94 &$-$29 03 15.3&1.92&$+$6.6&0.4&       620   \\
52&EI&17 45 38.00 &$-$29 05 34.9&0.13&$+$14.3&0.4&       42   \\
53&EG&17 45 38.01 &$-$29 03 39.0&0.14&$+$21.5&0.4&         45 \\
54&EI&17 45 38.03 &$-$29 05 34.8&0.53&$+$61.0&1.2&         170   \\
55&EH&17 45 38.07 &$-$29 04 27.6&0.39&$+$24.9&0.4&         130  \\
56&EH&17 45 38.18 &$-$29 04 29.5&0.13&$+$23.6&0.4&        42    \\
57&EI&17 45 38.20 &$-$29 05 32.2&0.07&$+$17.7&0.8&      23   \\
58&EH&17 45 38.33 &$-$29 05 13.7&0.19&$+$3.2&0.4&     61  \\
59&EF&17 45 38.37 &$-$29 02 56.6&0.25&$+$18.5&0.4&        81  \\

60&EG&17 45 38.39 &$-$29 03 32.9&0.14&$+$15.5&0.8&         45  &         M\\
61&AA&17 45 38.86 &$-$29 00 20.9&0.92&$+$43.6&0.4&      300  &         R\\
62&AA&17 45 39.03 &$-$29 00 16.9&0.06&$+$48.7&0.8&        19   &         R\\
63&DG&17 45 40.31 &$-$29 03 56.7&0.28&$+$19.8&0.4&       90   \\
64&DG&17 45 40.37 &$-$29 03 57.7&0.30&$+$16.0&0.4&       97    \\
65&DD&17 45 41.05 &$-$29 00 57.4&0.21&$+$15.1&0.4&        68   \\
66&DD&17 45 41.32 &$-$29 00 59.3&0.20&$+$18.1&1.2&      65  \\
67&DC&17 45 42.22 &$-$28 59 56.4&0.07&$+$48.3&0.4&      23 \\
68&DE&17 45 42.41 &$-$29 01 58.5&0.12&$+$24.9&0.4&      39   \\
69&DB&17 45 42.80 &$-$28 58 53.9&0.12&$+$47.9&0.4&    39  \\
70&DB&17 45 43.23 &$-$28 59 29.0&0.08&$+$49.6&0.4&     26    \\

71&DC&17 45 43.74 &$-$29 00 14.9&0.29&$+$53.8&0.4&         94   &         R\\
72&DC&17 45 43.78 &$-$29 00 18.0&0.14&$+$51.3&0.4&        45 \\
73&DD&17 45 43.80 &$-$29 00 29.5&0.63&$+$47.0&0.8&      200   \\
74&DC&17 45 43.85 &$-$29 00 25.8&0.12&$+$49.6&0.4&     39     \\
75&DC&17 45 43.88 &$-$29 00 16.6&0.22&$+$53.4&0.4&         71    \\
76&DC&17 45 43.90 &$-$29 00 23.0&0.10&$+$49.1&0.4&        32  \\
77&DC&17 45 43.93 &$-$29 00 19.8&4.80&$+$49.6&0.4&   1600  &                      R \\
78&DB&17 45 44.05 &$-$28 58 32.8&0.08&$+$42.3&0.4&          26   \\

79&DB&17 45 44.31 &$-$28 59 18.7&0.09&$+$49.6&0.4&           29  \\
80&DB&17 45 44.42 &$-$28 59 20.9&0.10&$+$42.3&0.4&           32  \\
81&DB&17 45 44.59 &$-$28 59 21.1&0.11&$+$49.1&0.4&        36    \\
82&CB&17 45 45.00 &$-$28 58 28.9&0.55&$-$10.0&0.4&           180    &         M\\
83&CB&17 45 45.06 &$-$28 58 27.6&1.37&$-$8.3&2.0&         440 &         M\\
84&CB&17 45 45.26 &$-$28 58 27.3&1.47&$-$12.9&0.8&       470   \\
85&CB&17 45 45.27 &$-$28 58 27.9&0.22&$-$1.9&0.4&         71  \\
86&CB&17 45 45.28 &$-$28 58 26.9&0.23&$-$7.0&0.4&         74   \\
87&CB&17 45 45.50 &$-$28 58 25.5&0.23&$+$9.2&0.4&            74          &         M\\

88&CB&17 45 45.50 &$-$28 58 25.0&0.16&$+$3.2&0.4&        52    \\
89&CB&17 45 46.63 &$-$28 58 34.7&0.33&$+$46.2&1.2&       110   &M\\
90&CB&17 45 47.66 &$-$28 58 41.7&0.19&$+$55.9&0.8&          61 \\
91&CC&17 45 47.75 &$-$28 59 50.6&0.07&$+$53.0&0.4&       23\\
92&CD&17 45 47.75 &$-$29 00 45.6&0.15&$+$57.2&0.4&        48   \\
93&CE&17 45 47.78 &$-$29 01 57.1&0.07&$+$39.8&0.4&       23    \\
94&CB&17 45 47.78 &$-$28 58 34.6&0.58&$+$39.8&0.8&      190   \\
95&CB&17 45 47.89 &$-$28 58 35.3&0.49&$+$52.5&1.2&       160    &M\\
96&CB&17 45 47.95 &$-$28 58 41.3&0.06&$+$49.6&0.4&         19   \\

97&CB&17 45 48.01 &$-$28 58 40.0&0.23&$+$44.0&0.4&        74   &M\\
98&CB&17 45 48.02 &$-$28 58 40.8&0.78&$+$31.3&0.8&       250    &M\\
99&CD&17 45 48.21 &$-$29 01 01.7&0.15&$+$55.9&0.4&          48   &M\\
100&CB&17 45 48.21 &$-$28 59 09.1&0.22&$+$50.4&0.4&           71  \\
101&CB&17 45 48.25 &$-$28 58 44.5&0.37&$+$43.6&0.4&       120    \\
102&CC&17 45 48.81 &$-$28 59 37.7&0.34&$+$41.5&0.4&     110     \\
103&CC&17 45 48.86 &$-$29 00 07.1&0.36&$+$35.5&0.4&   120   \\
104&CD&17 45 48.87 &$-$29 00 53.0&0.12&$+$41.1&0.4&         39    \\
105&CB&17 45 49.03 &$-$28 59 10.2&0.15&$+$42.3&0.8&       48    \\
106&BC&17 45 49.08 &$-$29 00 14.9&0.11&$+$27.4&0.4&        36    \\
107&BC&17 45 49.08 &$-$29 00 18.0&0.12&$+$36.4&0.4&      39    \\
108&BC&17 45 49.10 &$-$29 00 16.3&0.27&$+$57.6&0.4&      87   \\

109&BB&17 45 49.14 &$-$28 58 35.7&0.70&$+$51.7&0.4&     230   \\
110&BB&17 45 49.29 &$-$28 58 54.9&0.68&$+$48.7&1.2&       220    \\
111&BB&17 45 49.31 &$-$28 58 54.0&0.83&$+$43.2&0.8&       270  \\
112&BC&17 45 49.35 &$-$29 00 06.3&0.32&$+$49.6&0.4&       100     &M\\
113&BB&17 45 49.36 &$-$28 58 53.3&4.04&$+$44.9&0.8&  1300      &R\\
114&BB&17 45 49.39 &$-$28 58 52.9&2.09&$+$41.1&0.4&     670    &M\\
115&BB&17 45 49.40 &$-$28 58 54.1&0.18&$+$41.9&0.4&     58     &M\\
116&BB&17 45 49.44 &$-$28 58 52.3&1.04&$+$38.9&2.0&      340   &M\\
117&BB&17 45 49.50 &$-$28 58 52.6&0.21&$+$44.9&1.6&      68 &M\\

118&BB&17 45 49.53 &$-$28 58 52.3&0.49&$+$42.3&0.8&       160    \\
119&BD&17 45 49.54 &$-$29 00 31.3&0.09&$+$47.9&2.0&   29     &M\\
120&BB&17 45 49.56 &$-$28 59 00.6&5.91&$+$46.2&0.4&   1900 &M,R\\
121&BB&17 45 49.60 &$-$28 58 51.9&0.82&$+$40.6&0.4&     260     \\
122&BB&17 45 49.61 &$-$28 58 55.1&0.18&$+$42.3&0.4&     58   \\
123&BB&17 45 49.68 &$-$28 58 55.2&16.19&$+$42.3&0.4&     5200    &R\\
124&BC&17 45 49.69 &$-$28 59 59.3&0.14&$+$37.2&0.4&     45     \\
125&BB&17 45 49.69 &$-$28 58 54.2&0.96&$+$44.5&0.4&   310    \\
126&BD&17 45 49.73 &$-$29 00 40.4&0.44&$+$34.7&0.8&    140    &M\\
127&BD&17 45 49.73 &$-$29 00 39.7&0.42&$+$29.6&0.4&      140   \\

128&BB&17 45 49.74 &$-$28 59 04.5&0.22&$+$44.5&0.4&        71    \\
129&BC&17 45 49.77 &$-$29 00 01.9&1.19&$+$54.7&0.4&       380 &M\\
130&BB&17 45 49.87 &$-$28 59 03.6&0.11&$+$36.8&0.4&       36   \\
131&BC&17 45 49.88 &$-$29 00 22.3&0.14&$+$27.4&0.4&       45  \\
132&BB&17 45 49.89 &$-$28 59 04.5&9.95&$+$36.4&0.4&     3200  &R\\
133&BB&17 45 49.90 &$-$28 59 02.6&0.12&$+$46.6&0.4&    39      \\
134&BB&17 45 49.94 &$-$28 59 02.5&0.12&$+$41.9&0.8&       39   &M\\
135&BB&17 45 49.98 &$-$28 58 50.3&0.69&$+$58.5&0.4&       220    &M\\

136&BC&17 45 49.99 &$-$29 00 02.5&0.10&$+$47.4&0.4&        32   \\
137&BB&17 45 50.00 &$-$28 58 49.6&0.29&$+$53.4&0.4&         94    \\
138&BB&17 45 50.02 &$-$28 59 02.8&0.14&$+$47.0&0.4&        45   \\
139&BC&17 45 50.02 &$-$28 59 52.6&0.25&$+$34.3&0.4&     81 \\
140&BB&17 45 50.03 &$-$28 59 06.0&1.43&$+$38.1&0.4&        460    \\
141&BC&17 45 50.03 &$-$28 59 59.9&0.18&$+$55.1&0.8&    58    \\
142&BB&17 45 50.04 &$-$28 58 55.9&0.15&$+$42.3&0.4&     48  \\
143&BC&17 45 50.08 &$-$28 59 42.0&0.18&$+$40.6&0.4&        58  \\
144&BB&17 45 50.08 &$-$28 58 51.0&0.23&$+$54.7&0.4&     74   \\

145&BB&17 45 50.10 &$-$28 58 51.9&0.18&$+$43.2&0.4&       58     \\
146&BC&17 45 50.10 &$-$28 59 40.2&0.40&$+$39.4&0.4&        130   \\
147&BB&17 45 50.15 &$-$28 58 51.4&0.87&$+$52.1&0.4&     280   \\
148&BD&17 45 50.17 &$-$29 00 26.7&0.09&$+$49.6&0.4&          29  \\
149&BB&17 45 50.22 &$-$28 58 55.3&0.11&$+$41.9&0.4&    36   \\
150&BB&17 45 50.25 &$-$28 58 51.6&0.14&$+$50.8&0.4&         45    \\
151&BE&17 45 50.26 &$-$29 02 05.3&0.70&$+$67.8&0.8&    230     &M\\
152&BB&17 45 50.26 &$-$28 58 51.2&0.58&$+$42.3&0.8&     190     \\

153&BE&17 45 50.27 &$-$29 02 05.9&0.10&$+$65.7&0.4&       32    \\
154&BC&17 45 50.28 &$-$28 59 52.1&0.11&$+$40.2&0.4&       36     \\
155&BB&17 45 50.28 &$-$28 59 11.9&0.14&$+$41.9&0.4&        45   \\
156&BC&17 45 50.29 &$-$28 59 52.7&0.34&$+$37.2&0.4&       110 \\
157&BB&17 45 50.30 &$-$28 58 54.9&0.08&$+$36.4&0.4&        26    \\
158&BB&17 45 50.30 &$-$28 59 21.7&0.20&$+$51.7&0.4&     65    \\
159&BC&17 45 50.32 &$-$29 00 06.4&0.11&$+$52.1&0.4&        36   \\

160&BB&17 45 50.32 &$-$28 59 13.2&0.10&$+$31.7&2.4&      32    &M\\
161&BB&17 45 50.36 &$-$28 59 15.8&1.13&$+$34.7&0.4&     360     &M\\
162&BC&17 45 50.36 &$-$29 00 05.9&0.36&$+$47.9&0.4&         120  \\
163&BB&17 45 50.37 &$-$28 59 21.8&0.11&$+$48.7&0.4&      36     \\
164&BB&17 45 50.38 &$-$28 59 14.0&0.15&$+$32.6&0.4&         48     \\
165&BB&17 45 50.39 &$-$28 59 08.4&0.43&$+$42.3&0.4&        140   \\
166&BB&17 45 50.41 &$-$28 59 21.6&1.16&$+$47.0&0.4&    370     &M\\
167&BB&17 45 50.42 &$-$28 58 52.4&0.16&$+$43.2&0.4&      52  &M\\
168&BB&17 45 50.42 &$-$28 59 12.6&0.22&$+$28.7&0.8&         71  \\
169&BB&17 45 50.44 &$-$28 58 56.1&0.32&$+$50.0&0.4&       100  &M\\
170&BC&17 45 50.45 &$-$29 00 05.5&1.24&$+$49.1&1.2&       400   &R\\
171&BB&17 45 50.47 &$-$28 58 55.0&0.11&$+$41.9&0.4&       36   \\
172&BB&17 45 50.48 &$-$28 58 48.7&2.70&$+$58.9&0.8&     870  &M,R\\
173&BC&17 45 50.49 &$-$29 00 04.8&2.35&$+$47.9&2.4&           760   \\

174&BB&17 45 50.49 &$-$28 59 21.2&0.12&$+$36.4&0.4&      39  \\
175&BB&17 45 50.49 &$-$28 59 07.6&2.45&$+$41.5&2.0&   790      &M,R\\
176&BB&17 45 50.50 &$-$28 58 52.5&0.20&$+$34.7&0.4&     65     \\
177&BC&17 45 50.52 &$-$29 00 06.0&0.08&$+$47.9&0.4&    26     &M\\
178&BB&17 45 50.52 &$-$28 59 11.5&0.11&$+$33.0&1.2&   36     \\
179&BB&17 45 50.56 &$-$28 59 08.5&0.79&$+$33.8&0.4&       250\\
180&BB&17 45 50.58 &$-$28 59 12.6&0.44&$+$36.0&0.4&         140 \\

181&BB&17 45 50.59 &$-$28 59 09.0&0.36&$+$37.2&2.8&        120  &M\\
182&BB&17 45 50.60 &$-$28 59 10.0&0.23&$+$23.2&1.2&          74 &M\\
183&BB&17 45 50.62 &$-$28 59 20.8&0.11&$+$41.9&0.4&   36 \\
184&BB&17 45 50.64 &$-$28 59 07.9&0.13&$+$27.4&0.4&      42  \\
185&BB&17 45 50.65 &$-$28 59 07.5&0.12&$+$35.1&0.4&       39   &M\\
186&BC&17 45 50.65 &$-$29 00 05.8&0.14&$+$46.6&0.4&       45   \\
187&BB&17 45 50.66 &$-$28 59 08.8&7.48&$+$27.0&0.4&    2400   &M,R\\
188&BC&17 45 50.68 &$-$28 59 53.9&0.10&$+$28.3&0.8&       32    &M\\
189&BD&17 45 50.68 &$-$29 00 35.4&0.11&$+$39.4&0.8&         36  &M\\
190&BB&17 45 50.68 &$-$28 59 11.4&0.12&$+$32.6&0.8&         39    \\
191&BB&17 45 50.69 &$-$28 58 55.1&0.21&$+$41.9&0.4&        68   \\
192&BB&17 45 50.75 &$-$28 58 58.1&0.10&$+$46.6&0.4&        32     \\

193&BB&17 45 50.75 &$-$28 59 06.7&0.17&$+$50.0&0.8&        55     &M\\
194&BB&17 45 50.77 &$-$28 58 37.9&0.15&$+$47.0&0.4&      48    \\
195&BE&17 45 50.78 &$-$29 01 45.9&0.76&$+$71.7&1.2&   250   &M\\
196&BB&17 45 50.83 &$-$28 58 56.2&1.12&$+$55.5&0.4&         360   \\
197&BB&17 45 50.85 &$-$28 58 59.8&0.45&$+$30.4&0.4&      150   &M\\
198&BE&17 45 50.88 &$-$29 01 32.6&0.14&$+$46.2&0.4&       45   \\
199&BC&17 45 50.88 &$-$28 59 41.7&0.13&$+$40.6&0.4&         42     \\
200&BB&17 45 50.88 &$-$28 58 47.7&0.12&$+$27.0&0.4&         39   \\
201&BB&17 45 50.89 &$-$28 58 54.2&0.09&$+$47.4&0.4&         29   \\
202&BC&17 45 50.89 &$-$29 00 04.9&0.23&$+$47.9&0.4&       74  \\

203&BB&17 45 50.90 &$-$28 59 04.4&0.09&$+$36.4&0.4&         29    \\
204&BC&17 45 50.92 &$-$28 59 58.6&0.75&$+$51.3&2.4&   240   &M\\
205&BC&17 45 51.01 &$-$29 00 12.6&0.13&$+$28.7&1.2&       42\\
206&BB&17 45 51.02 &$-$28 59 19.3&0.13&$+$44.5&1.2&       42  \\
207&BC&17 45 51.05 &$-$29 00 12.7&0.09&$+$33.0&0.4&         29   \\
208&BC&17 45 51.07 &$-$29 00 12.0&0.09&$+$29.6&1.6&         29  &M\\
209&BC&17 45 51.12 &$-$28 59 43.6&0.70&$+$37.7&1.6&    230   \\
210&BC&17 45 51.14 &$-$29 00 11.1&0.10&$+$39.4&0.4&        32  \\

211&BC&17 45 51.14 &$-$29 00 12.3&0.07&$+$35.1&0.8&         23    &M\\
212&BC&17 45 51.15 &$-$29 00 11.7&0.54&$+$37.2&0.8&      170    \\
213&BC&17 45 51.16 &$-$29 00 10.8&0.08&$+$37.2&0.4&        26   \\
214&BC&17 45 51.16 &$-$29 00 12.4&0.16&$+$34.7&1.2&       52    &M\\
215&BC&17 45 51.17 &$-$29 00 13.8&0.13&$+$34.7&0.4&          42  \\
216&BB&17 45 51.17 &$-$28 58 52.0&0.12&$+$47.4&1.2&         39   \\
217&BC&17 45 51.18 &$-$29 00 15.5&0.95&$+$33.4&0.8&     310   &M\\
218&BB&17 45 51.19 &$-$28 59 17.1&0.15&$+$38.9&0.4&       48   \\
219&BC&17 45 51.20 &$-$29 00 12.2&0.17&$+$39.8&0.4&      55      &M\\
220&BC&17 45 51.20 &$-$29 00 10.5&0.31&$+$38.1&0.8&      100    &M\\
221&BC&17 45 51.21 &$-$29 00 14.7&0.09&$+$34.3&0.8&          29  &M\\
222&BC&17 45 51.21 &$-$29 00 11.9&0.25&$+$42.8&1.2&      81     &M\\
223&BB&17 45 51.22 &$-$28 58 51.6&0.15&$+$47.0&0.8&       48   &M\\

224&BC&17 45 51.24 &$-$29 00 16.1&0.27&$+$33.8&0.4&         87    &M\\
225&BB&17 45 51.24 &$-$28 59 10.1&0.15&$+$38.1&0.4&       48    \\
226&BC&17 45 51.25 &$-$29 00 13.1&0.51&$+$36.8&1.6&      170      \\
227&BC&17 45 51.26 &$-$28 59 59.5&0.07&$+$36.4&0.4&        23     \\
228&BC&17 45 51.27 &$-$29 00 12.0&0.94&$+$39.4&1.2&     300     &M\\
229&BC&17 45 51.30 &$-$29 00 14.7&0.25&$+$36.4&0.4&       81  &M\\
230&BB&17 45 51.32 &$-$28 59 08.0&0.26&$+$44.9&0.4&         84    \\
231&BC&17 45 51.32 &$-$29 00 11.9&0.97&$+$34.3&0.8&      310     &M\\
232&BC&17 45 51.33 &$-$29 00 11.4&0.54&$+$29.1&0.4&    170     \\
233&BB&17 45 51.34 &$-$28 58 47.4&0.11&$+$46.6&0.4&        36   \\
234&BC&17 45 51.36 &$-$29 00 14.5&0.11&$+$35.1&1.2&      36  &M\\
235&BC&17 45 51.36 &$-$29 00 12.4&0.18&$+$40.6&0.4&        58  \\
236&BC&17 45 51.39 &$-$29 00 14.3&0.11&$+$33.4&0.4&     36     \\
237&BB&17 45 51.40 &$-$28 59 11.7&0.19&$+$41.9&0.4&    61 \\
238&BB&17 45 51.42 &$-$28 59 10.2&1.19&$+$43.6&0.4&    380 &M,R\\
239&BB&17 45 51.43 &$-$28 59 13.3&0.16&$+$49.6&1.6&     52     &M\\
240&BB&17 45 51.43 &$-$28 59 02.1&0.12&$+$42.3&0.4&     39   \\
241&BC&17 45 51.44 &$-$29 00 10.0&0.64&$+$30.9&1.2&    210     \\
242&BB&17 45 51.44 &$-$28 58 53.6&0.09&$+$46.2&0.4&       29 \\
243&BB&17 45 51.45 &$-$28 59 09.7&0.35&$+$40.6&0.8&     110\\
244&BB&17 45 51.47 &$-$28 58 54.9&0.07&$+$41.9&0.4&         23    \\
245&BB&17 45 51.49 &$-$28 59 13.4&0.25&$+$48.3&0.8&           81  \\
246&BC&17 45 51.50 &$-$29 00 10.0&0.28&$+$27.4&0.8&          90    \\
247&BB&17 45 51.53 &$-$28 58 53.9&0.53&$+$45.3&2.4&        170    &M\\

248&BC&17 45 51.53 &$-$29 00 09.3&0.18&$+$29.1&0.4&         58     \\
249&BC&17 45 51.54 &$-$29 00 10.3&0.07&$+$37.7&0.8&      23    \\
250&BC&17 45 51.54 &$-$29 00 12.0&0.10&$+$32.1&0.4&      32    \\
251&BC&17 45 51.56 &$-$29 00 10.1&0.09&$+$28.7&0.8&       29 \\
252&BB&17 45 51.56 &$-$28 58 53.7&2.03&$+$46.2&1.6&      650     &R\\
253&BC&17 45 51.59 &$-$28 59 37.2&0.33&$+$38.1&0.4&       110    \\
254&BB&17 45 51.62 &$-$28 58 28.6&0.26&$+$54.2&0.4&     84     \\
255&BB&17 45 51.62 &$-$28 59 07.8&0.49&$+$41.9&0.4&    160     \\
256&BB&17 45 51.62 &$-$28 58 26.7&0.74&$+$53.0&0.8&        240 & R\\
257&BC&17 45 51.64 &$-$29 00 06.8&0.14&$+$40.6&0.4&    45   \\
258&BC&17 45 51.66 &$-$29 00 09.8&0.08&$+$27.4&0.4&       26   \\
259&BB&17 45 51.68 &$-$28 58 51.2&0.71&$+$60.2&0.4&        230    &M\\
260&BB&17 45 51.70 &$-$28 58 40.6&0.19&$+$46.6&0.4&        61   \\
261&BB&17 45 51.70 &$-$28 58 50.8&0.99&$+$56.8&0.4&   320    \\
262&BB&17 45 51.71 &$-$28 58 56.6&0.18&$+$44.9&0.8&         58  &M\\

263&BB&17 45 51.72 &$-$28 58 53.0&1.16&$+$58.5&0.8&      370    &M\\
264&BB&17 45 51.75 &$-$28 58 51.8&0.19&$+$51.7&0.4&       61    \\
265&BB&17 45 51.76 &$-$28 58 52.9&0.16&$+$53.8&1.2&    52   \\
266&BB&17 45 51.78 &$-$28 58 52.2&0.61&$+$49.1&0.8&         200  \\
267&BB&17 45 51.88 &$-$28 58 52.9&11.90&$+$47.0&0.4&     3800   &R\\
268&BB&17 45 51.90 &$-$28 58 47.1&0.12&$+$46.6&0.4&      39  &M\\
269&BB&17 45 51.94 &$-$28 58 53.4&0.94&$+$49.6&0.8&     300      &M\\

270&BB&17 45 51.97 &$-$28 58 54.6&0.11&$+$51.7&2.0&         36   &M\\
271&BC&17 45 51.98 &$-$28 59 32.9&0.24&$+$41.1&0.4&          77  \\
272&BB&17 45 51.99 &$-$28 58 54.4&0.24&$+$51.7&0.8&         77 &M\\
273&BB&17 45 52.00 &$-$28 58 53.1&0.10&$+$49.1&0.8&     32    &M\\
274&BB&17 45 52.02 &$-$28 58 53.9&0.22&$+$53.4&0.4&       71     \\
275&BB&17 45 52.03 &$-$28 58 54.7&0.47&$+$50.8&0.8&   150  &M\\
276&BB&17 45 52.03 &$-$28 58 53.5&0.29&$+$48.3&1.2&     94   &M\\
277&BB&17 45 52.04 &$-$28 58 55.8&0.18&$+$38.9&0.4&       58   \\

278&BB&17 45 52.05 &$-$28 58 50.3&0.45&$+$46.2&0.4&     150    &M\\
279&BB&17 45 52.13 &$-$28 58 55.5&2.64&$+$45.3&0.8&      850    &M,R\\
280&BB&17 45 52.16 &$-$28 58 22.4&1.40&$+$54.7&1.2&   450    &R\\
281&BB&17 45 52.19 &$-$28 58 55.3&0.13&$+$47.9&0.4&    42    \\
282&BB&17 45 52.36 &$-$28 58 58.1&0.72&$+$45.3&1.2&     230   \\
283&BB&17 45 52.36 &$-$28 58 49.2&0.13&$+$41.9&1.6&      42    &M\\
284&BB&17 45 52.39 &$-$28 58 58.7&0.33&$+$45.3&0.8&       110    &M\\
285&BB&17 45 52.42 &$-$28 58 56.3&0.10&$+$44.9&0.4&       32     \\
286&BB&17 45 52.43 &$-$28 58 58.2&3.32&$+$43.2&0.8& 1070&M,R\\
287&BB&17 45 52.46 &$-$28 58 54.9&0.17&$+$42.3&0.8&     55 &M,R\\

288&BC&17 45 52.51 &$-$29 00 09.7&0.36&$+$68.7&0.8&       116    \\
289&BB&17 45 52.52 &$-$28 58 52.4&0.36&$+$37.7&0.4&         116   \\
290&BB&17 45 52.55 &$-$28 58 56.8&2.48&$+$45.3&0.4&           800   &R\\
291&BB&17 45 52.56 &$-$28 58 48.5&0.23&$+$48.7&0.4&         74  &M\\
292&BB&17 45 52.62 &$-$28 58 58.5&0.20&$+$44.0&0.4&        65   \\
293&BB&17 45 52.75 &$-$28 59 00.1&0.11&$+$31.3&0.4&          36   \\

294&BB&17 45 52.83 &$-$28 58 52.4&0.16&$+$46.2&0.4&          52  \\
295&BB&17 45 53.03 &$-$28 58 59.6&0.88&$+$48.3&0.4&     280  \\
296&BB&17 45 53.06 &$-$28 58 57.7&0.15&$+$46.6&0.4&  48 \\
297&BB&17 45 53.32 &$-$28 59 02.0&0.17&$+$43.6&0.4&      55    \\
298&BB&17 45 53.44 &$-$28 58 58.6&0.14&$+$30.0&0.4&      45  \\
299&AB&17 45 53.72 &$-$28 59 07.8&0.11&$+$47.0&0.4&       36  \\
300&BB&17 45 53.73 &$-$28 59 06.5&0.38&$+$42.3&0.8&      120 \\
301&AB&17 45 53.74 &$-$28 59 07.2&0.24&$+$41.5&0.4&       77     &M\\
302&AB&17 45 53.74 &$-$28 59 07.1&0.45&$+$41.5&0.4&        150 \\

303&BB&17 45 53.75 &$-$28 59 18.0&0.36&$+$29.1&0.4&      116   \\
304&AB&17 45 53.76 &$-$28 58 42.9&0.10&$+$46.6&0.4&         32  \\
305&AB&17 45 53.94 &$-$28 59 19.6&0.10&$+$42.3&0.4&     32    \\
306&AB&17 45 54.12 &$-$28 59 11.7&0.09&$+$38.9&0.4&      29    \\
307&AB&17 45 54.48 &$-$28 59 06.5&0.11&$+$43.2&0.4&          36 \\
308&AB&17 45 54.97 &$-$28 59 20.4&0.07&$+$24.5&0.8&      23  \\
309&AB&17 45 55.01 &$-$28 59 20.5&0.12&$+$18.9&0.8&         39    \\
310&AB&17 45 55.03 &$-$28 59 20.2&0.15&$+$31.7&0.4&     48   \\
311&AB&17 45 55.04 &$-$28 59 21.0&0.18&$+$22.3&0.4&      58  \\
312&AB&17 45 55.05 &$-$28 58 48.0&0.11&$+$37.2&0.4&      36    \\
313&AB&17 45 55.06 &$-$28 59 20.8&0.39&$+$16.4&0.8&      130  \\
314&AB&17 45 55.07 &$-$28 59 03.3&0.07&$+$42.3&0.4&     23\\
315&AB&17 45 55.44 &$-$28 59 00.2&0.06&$+$28.3&0.4&        19   \\
316&AB&17 45 55.93 &$-$28 59 13.1&0.31&$+$13.4&2.8&         100    &M\\
317&AB&17 45 56.10 &$-$28 58 45.0&0.38&$+$27.9&0.4&     120    &M\\
318&AB&17 45 57.15 &$-$28 59 16.2&0.54&$+$10.0&0.4&       170  \\

\label{masers}
\end{longtable*}
\footnotesize{Notes. (M): Multiple spectral features detected at the same position (above 10$\sigma$).  The spectral parameters of the brightest peak are listed in this table. (R): This 44 GHz maser emission was previously detected by Yusef-Zadeh et al. (2008) and/or Pihlstr{\"o}m et al. (2011).  The brightness temperatures, $T_b$, listen in column 8 are lower limits.}

\begin{figure*}[tbh]
\centering
\includegraphics[angle=0,scale=.3]{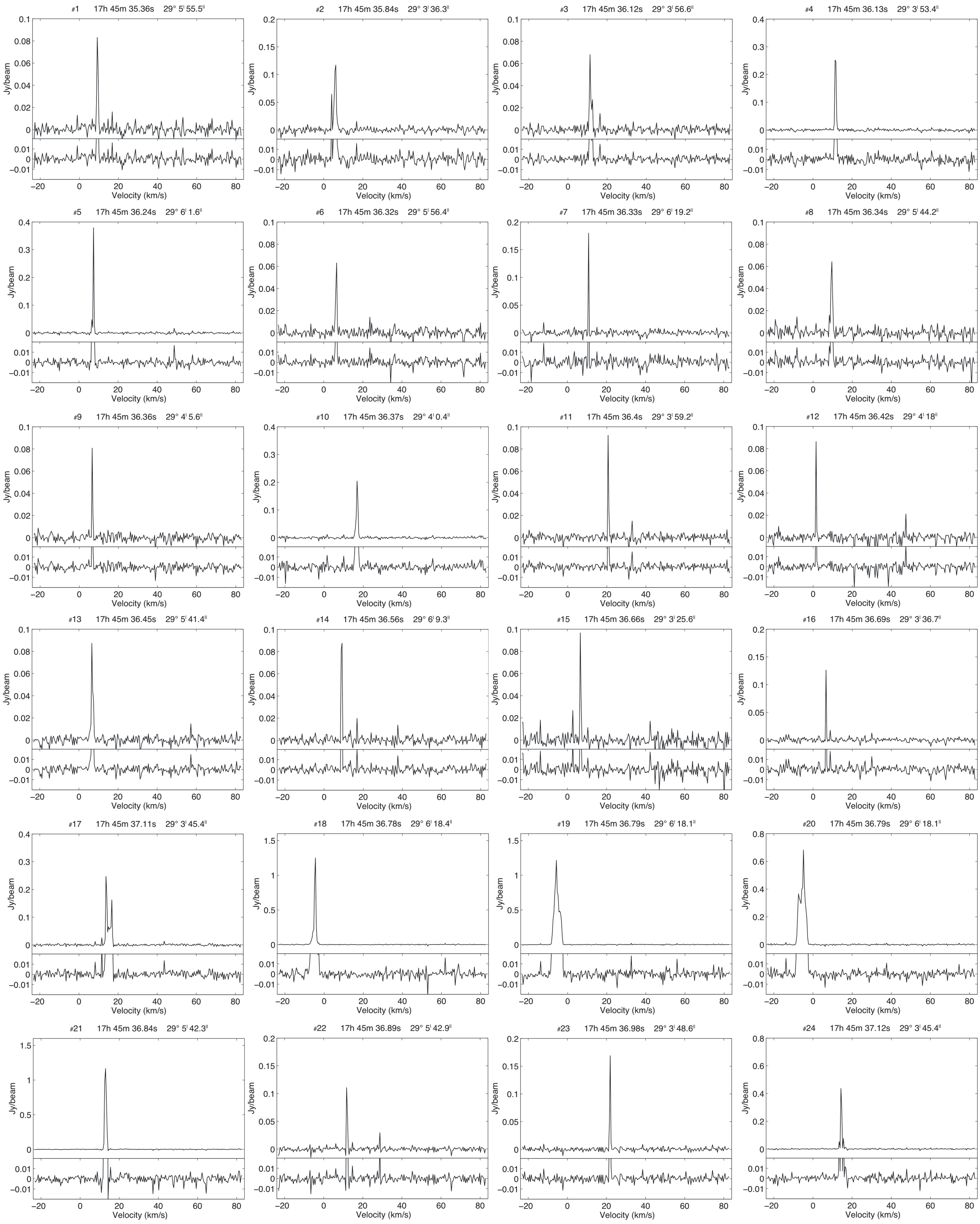}
\caption{Example 44 GHz CH$_3$OH maser spectral profiles as a function of LSR line of sight velocity.  The bottom row is a fixed scale zoom on the spectral baseline to more predominately show the rms noise and possible multiple spectral lines that may be invisible in the top row auto-scaled spectra.  The labels in the header of each spectra are the entry number in Table 1 and the J2000 position of the brightest peak.}
\label{spectra1}
\end{figure*}

\addtocounter{figure}{-1}
\begin{figure*}[tbh]
\centering
\includegraphics[angle=0,scale=.3]{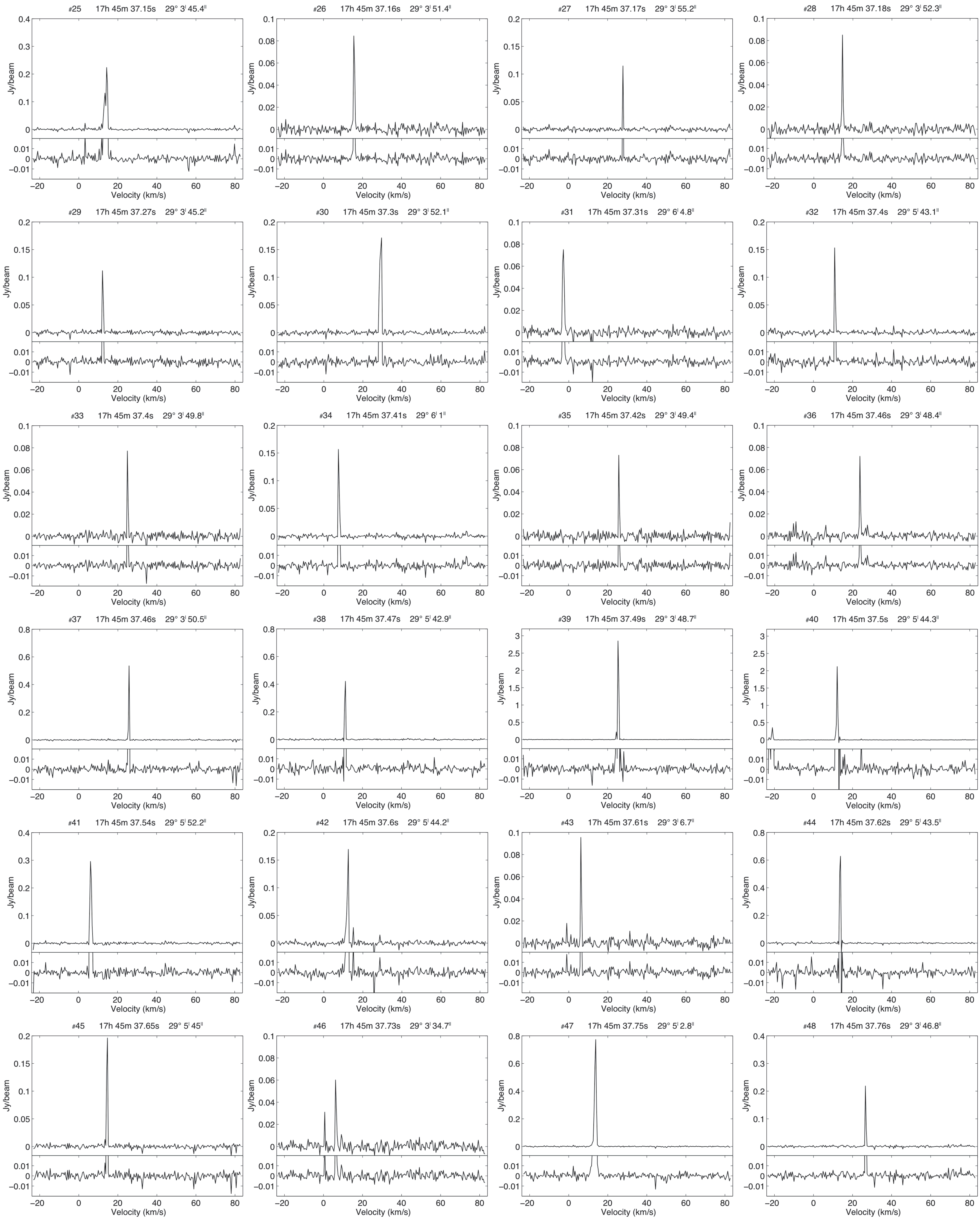}
\caption{Example 44 GHz CH$_3$OH maser spectral profiles continued...}
\label{spectra2}
\end{figure*}

\addtocounter{figure}{-1}
\begin{figure*}[tbh]
\centering
\includegraphics[angle=0,scale=.3]{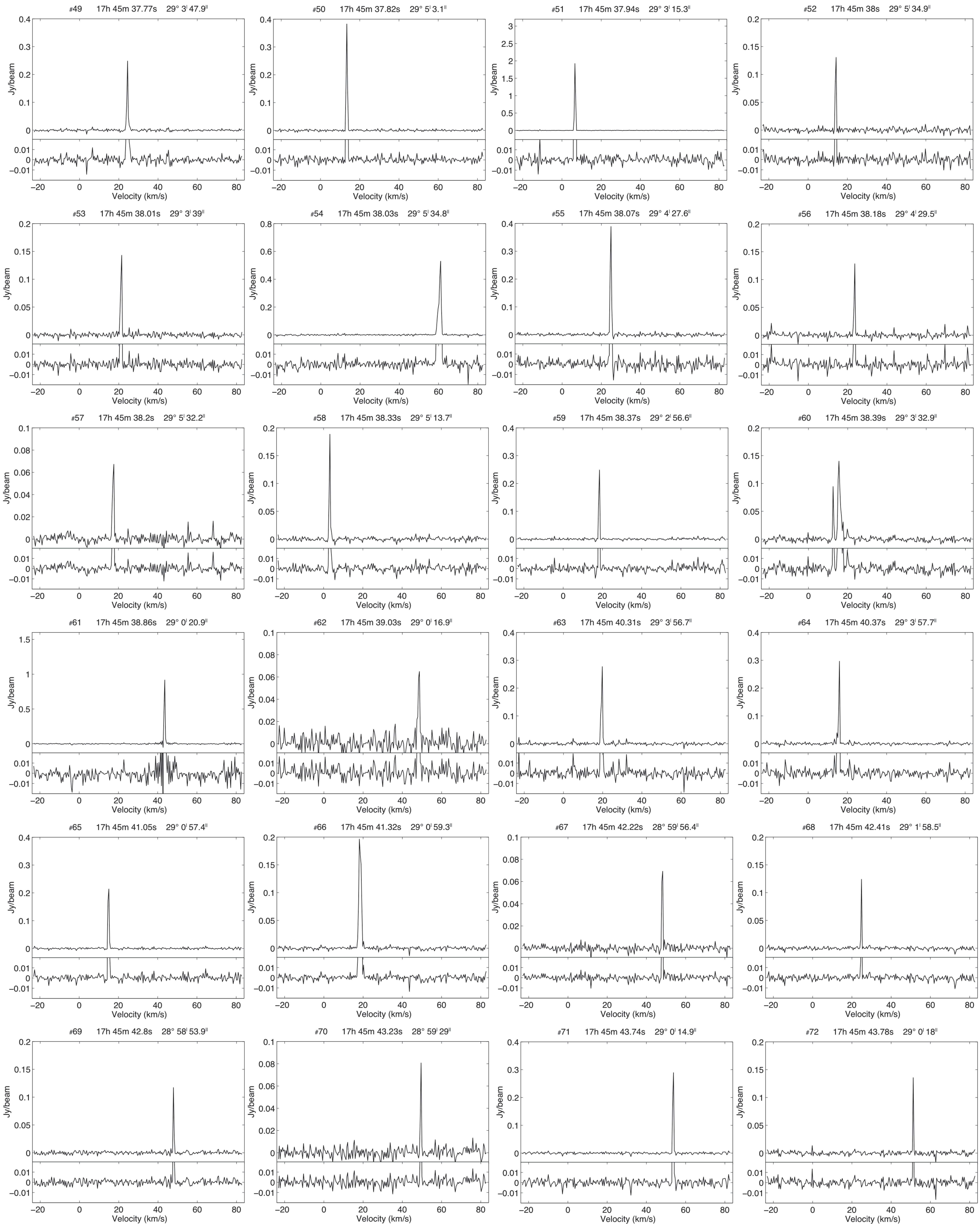}
\caption{Example 44 GHz CH$_3$OH maser spectral profiles continued...}
\label{spectra3}
\end{figure*}

\addtocounter{figure}{-1}
\begin{figure*}[tbh]
\centering
\includegraphics[angle=0,scale=.3]{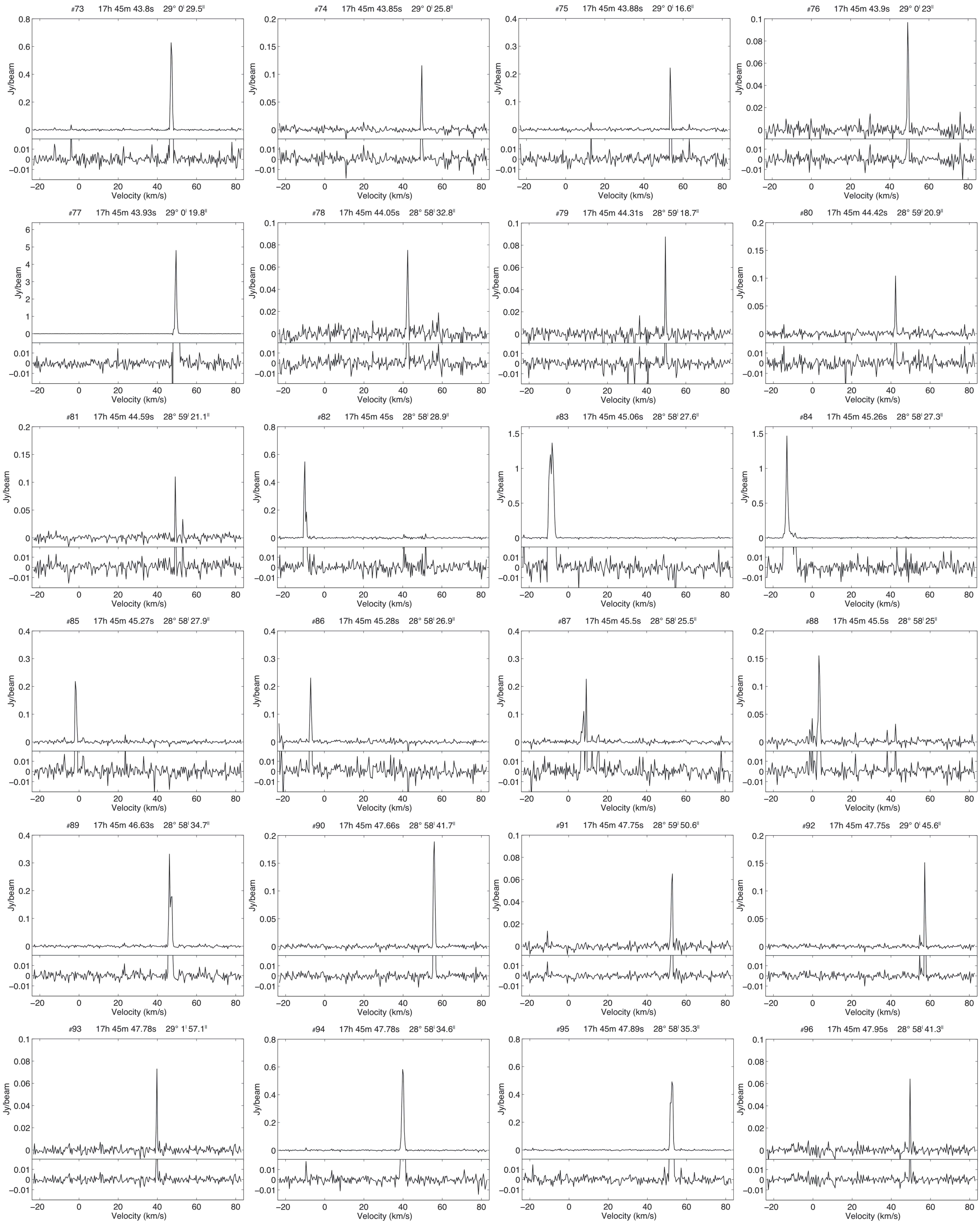}
\caption{Example 44 GHz CH$_3$OH maser spectral profiles continued...}
\label{spectra4}
\end{figure*}

\addtocounter{figure}{-1}
\begin{figure*}[tbh]
\centering
\includegraphics[angle=0,scale=.3]{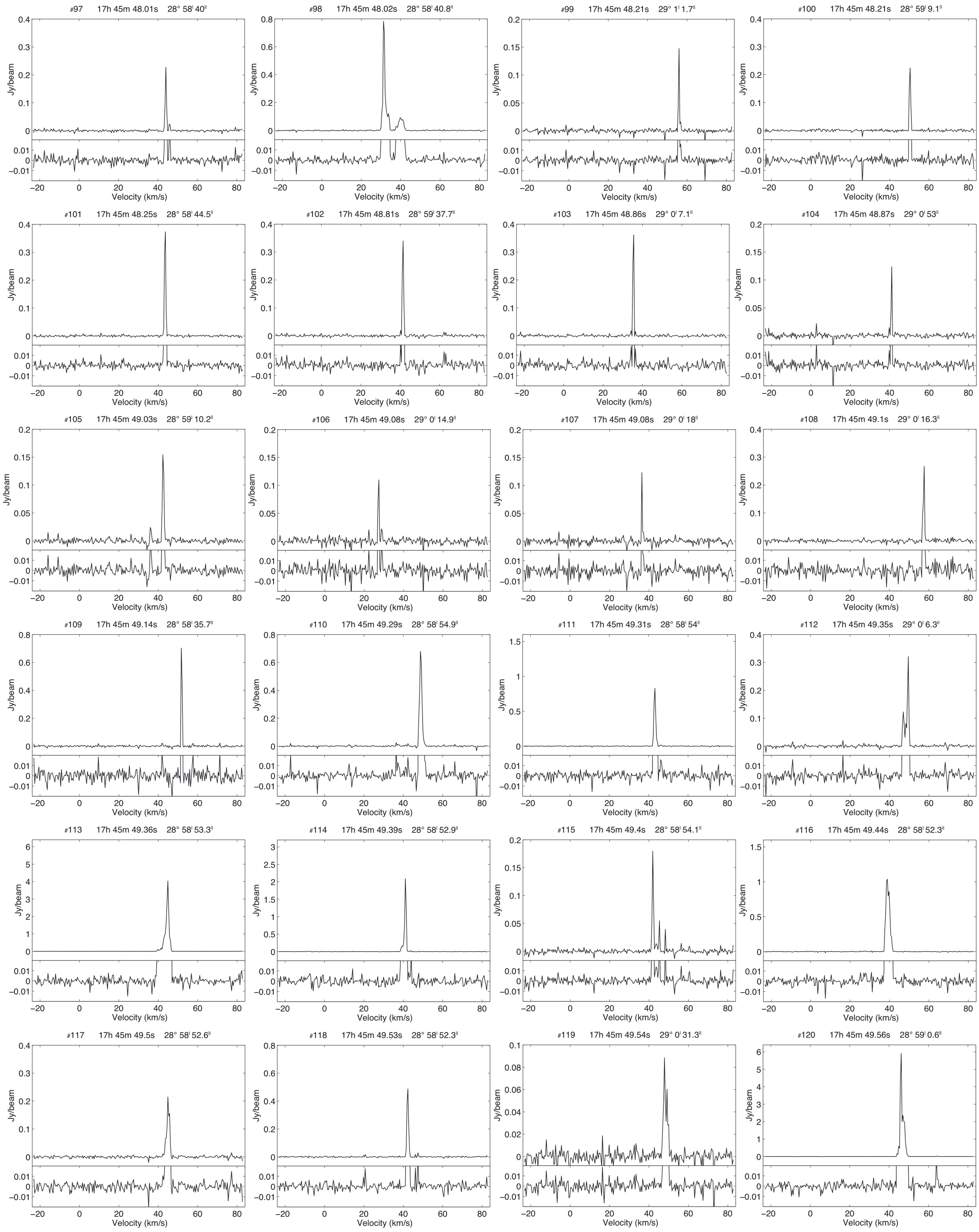}
\caption{Example 44 GHz CH$_3$OH maser spectral profiles continued...}
\label{spectra5}
\end{figure*}

\addtocounter{figure}{-1}
\begin{figure*}[tbh]
\centering
\includegraphics[angle=0,scale=.3]{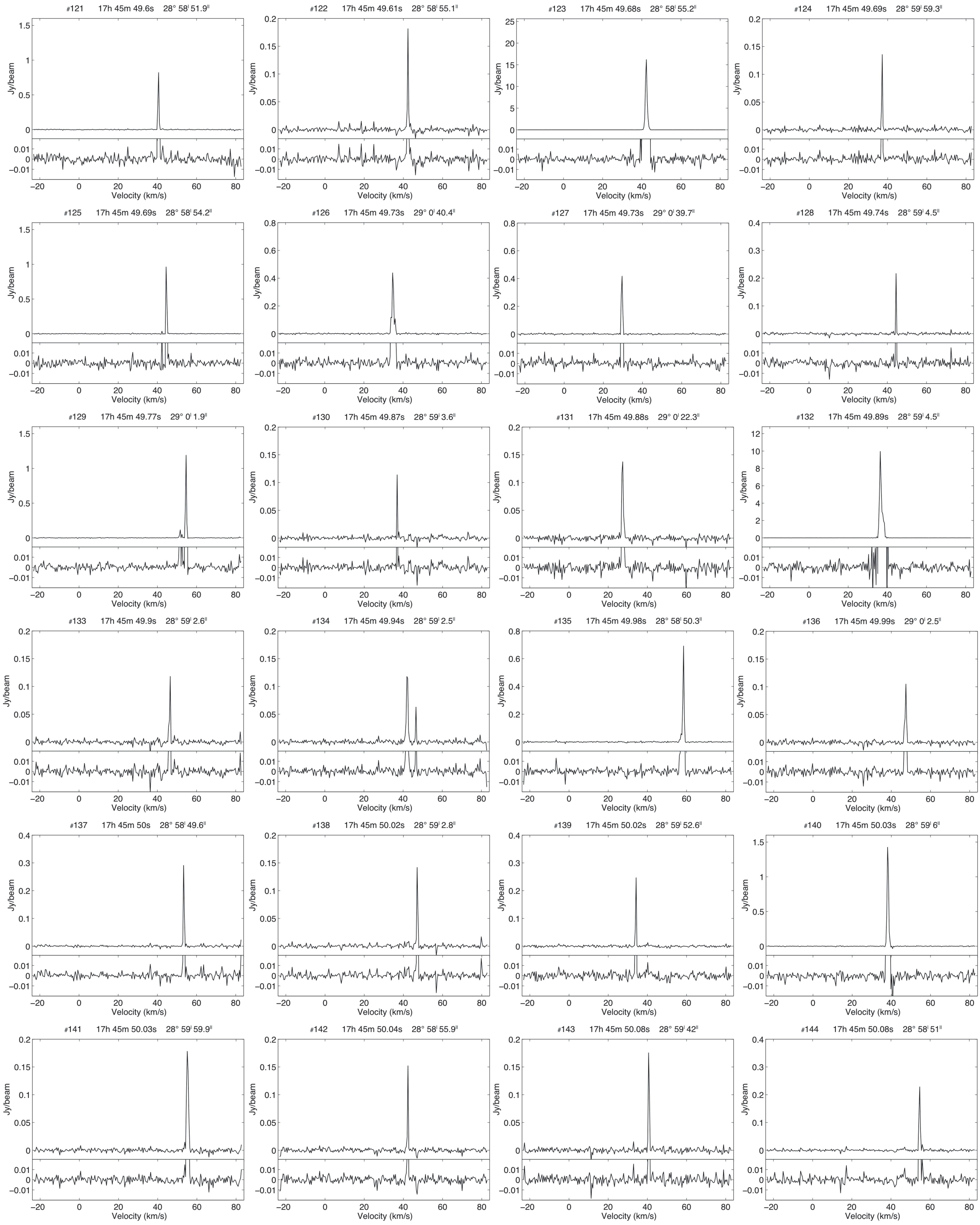}
\caption{Example 44 GHz CH$_3$OH maser spectral profiles continued...}
\label{spectra6}
\end{figure*}

\addtocounter{figure}{-1}
\begin{figure*}[tbh]
\centering
\includegraphics[angle=0,scale=.3]{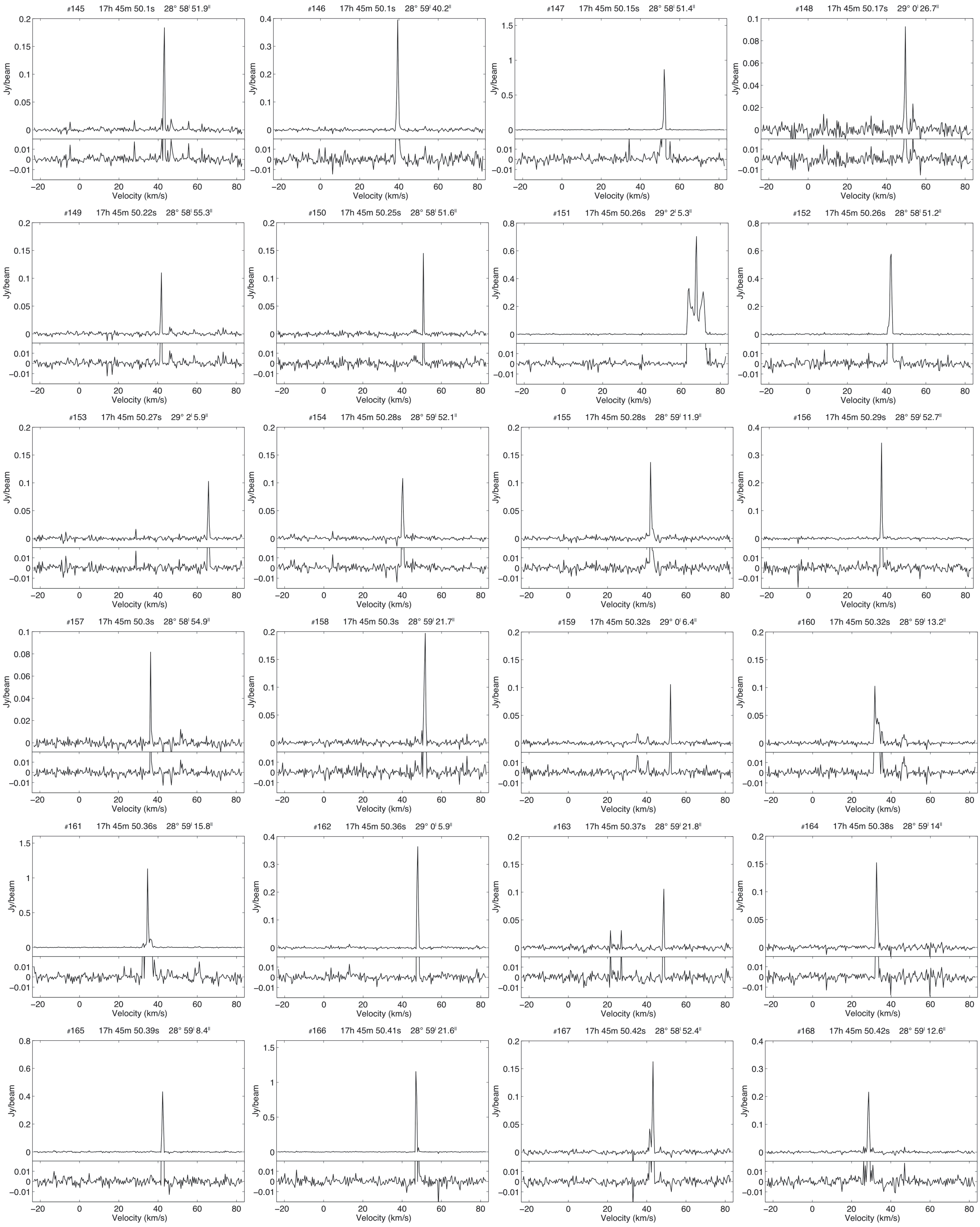}
\caption{Example 44 GHz CH$_3$OH maser spectral profiles continued...}
\label{spectra7}
\end{figure*}

%
%
%
%
%
%

\end{document}